\documentclass[twocolumn,english,british,prb,showpacs,floatfix,superscriptaddress]{revtex4-1}
\usepackage[T1]{fontenc}
\usepackage[latin9]{inputenc}
\usepackage{color}
\usepackage{amsmath}
\usepackage{amssymb}
\usepackage{graphicx}
\usepackage{epstopdf} 

\makeatletter

\newcommand*\LyXThinSpace{\,\hspace{0pt}}
\providecommand{\tabularnewline}{\\}

\usepackage{graphicx}
\usepackage{epstopdf} 
\makeatother

\usepackage{babel}

\begin{document}

\title{Nature of the many-body excitations in a quantum wire: theory and experiment}

\author{O. Tsyplyatyev}

\affiliation{Institut f{\"u}r Theoretische Physik, Universit{\"a}t Frankfurt,
Max-von-Laue Strasse 1, 60438 Frankfurt, Germany}

\author{A. J. Schofield}

\affiliation{School of Physics and Astronomy, University of Birmingham, Birmingham,
B15 2TT, UK}

\author{Y. Jin}

\affiliation{Cavendish Laboratory, University of Cambridge, J J Thomson Avenue,
Cambridge, CB3 0HE, UK}

\author{M. Moreno}

\affiliation{Cavendish Laboratory, University of Cambridge, J J Thomson Avenue,
Cambridge, CB3 0HE, UK}

\author{W. K. Tan}

\affiliation{Cavendish Laboratory, University of Cambridge, J J Thomson Avenue,
Cambridge, CB3 0HE, UK}

\author{A. S. Anirban}

\affiliation{Cavendish Laboratory, University of Cambridge, J J Thomson Avenue,
Cambridge, CB3 0HE, UK}

\author{C. J. B. Ford}

\affiliation{Cavendish Laboratory, University of Cambridge, J J Thomson Avenue,
Cambridge, CB3 0HE, UK}

\author{J. P. Griffiths}

\affiliation{Cavendish Laboratory, University of Cambridge, J J Thomson Avenue,
Cambridge, CB3 0HE, UK}

\author{I. Farrer}

\affiliation{Cavendish Laboratory, University of Cambridge, J J Thomson Avenue,
Cambridge, CB3 0HE, UK}

\author{G. A. C. Jones}

\affiliation{Cavendish Laboratory, University of Cambridge, J J Thomson Avenue,
Cambridge, CB3 0HE, UK}

\author{D. A. Ritchie}

\affiliation{Cavendish Laboratory, University of Cambridge, J J Thomson Avenue,
Cambridge, CB3 0HE, UK}

\date{\today}
\begin{abstract}
The natural excitations of an interacting one-dimensional system at low
energy are hydrodynamic modes of Luttinger liquid, protected by the
Lorentz invariance of the linear dispersion. We show that beyond low
energies, where quadratic dispersion reduces the symmetry to Galilean,
the main character of the many-body excitations changes into a hierarchy:
calculations of dynamic correlation functions for fermions (without
spin) show that the spectral weights of the excitations are proportional
to powers of $\mathcal{R}^{2}/L^{2}$, where $\mathcal{R}$ is a length-scale
related to interactions and $L$ is the system length. Thus only small
numbers of excitations carry the principal spectral power in representative
regions on the energy-momentum planes. We have analysed the spectral
function in detail and have shown that the first-level (strongest)
excitations form a mode with parabolic dispersion, like that of a
renormalised single particle. The second-level excitations produce
a singular power-law line shape to the first-level mode and multiple
power-laws at the spectral edge. We have illustrated crossover to
Luttinger liquid at low energy by calculating the local density of
state through all energy scales: from linear to non-linear, and to
above the chemical potential energies. In order to test this model, we have carried out experiments to measure momentum-resolved tunnelling of electrons (fermions with spin) from/to a wire formed within a GaAs heterostructure. We observe well-resolved spin-charge separation at low energy with appreciable interaction strength and only a parabolic dispersion of the first-level mode at higher energies. We find structure resembling the second-level excitations, which dies away rapidly at high momentum in line with the theoretical predictions here.
\end{abstract}

\maketitle

\section{Introduction}

Predicting the behaviour of interacting electrons is a significant open problem. Most progress to date has been made at low
energies where linearisation of the single-particle dispersion led
to construction of Fermi\cite{NozieresBook} and,
in one dimension, to Luttinger-liquid theories\cite{GiamarchiBook} in which the natural excitations
are fermionic quasiparticles and hydrodynamic modes respectively.
The only significant progress beyond the linear approximation has
been achieved via the heavy impurity model, for Fermi\cite{Nozieres69,Nozieres69_2,NozieresDeDominicis}
and Luttinger\cite{GlazmanReview12} liquids, showcasing
threshold singularities drastically different from the low energy
behaviour. In this paper we investigate one-dimensional (1D) fermions beyond
the linear approximation where the natural many-body excitations form
a hierarchical structure,\cite{TSFetal15} in sharp contrast with
the Fermi quasiparticles or hydrodynamic modes. We obtain the dynamical structure factor, in addition to the already known spectral function, and construct an inductive proof for calculating the form factors that are necessary for the dynamical response functions of the spinless fermion model. Experimentally, we demonstrate control over the interaction energy a 1D wire manifested as a change of the ratio of the charge and spin velocities at low energy scales. We find a new structure resembling the second-level excitations, which dies rapidly away from the first-level mode in a manner consistent with a power law.
\begin{figure}[h!]
{\centering\includegraphics[width=0.87\columnwidth]{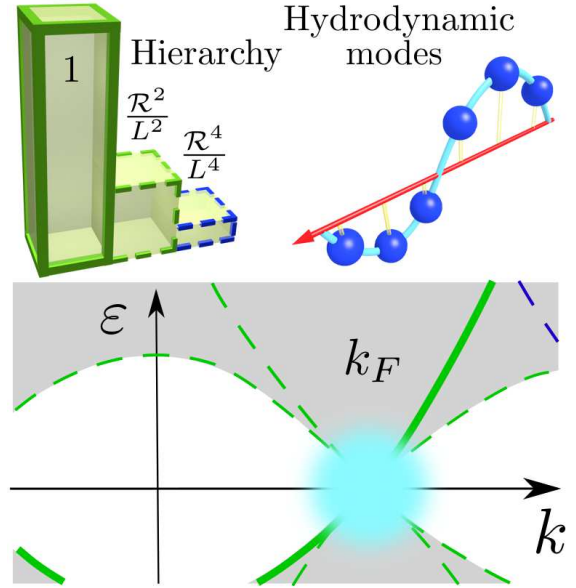} }
\protect\caption{\label{fig:fig_main} Regions of the energy-momentum plane dominated by two different principal regimes of the system (bottom): hydrodynamic modes of the Luttinger liquid (top-right) at low energies (marked with cyan colour in the bottom panel) and the hierarchy of modes (top-left) in the rest of the plane.
}
\end{figure}

We analyse theoretically the dynamic response functions\textendash that
probe the many-body excitations\textendash for spinless fermions with
short-range interactions. Our approach is exact diagonalisation via
Bethe ansatz methods: the eigenenergies are evaluated in the coordinate
representation and form factors\textendash for the corresponding eigenstates\textendash are
derived in the algebraic representation, via Slavnov's formula.\cite{Slavnov89}
On the microscopic level the excitations arrange themselves into a
hierarchy via their spectral weights\textendash given by the form
factors\textendash with different powers of $\mathcal{R}^{2}/L^{2}$,
where $\mathcal{R}$ is the particle-particle interaction radius and
$L$ is the length of the system. As a result only small numbers of
states\textendash out of an exponentially large Fock space of the
many-body system\textendash carry the principal spectral power in
representative regions on the energy-momentum plane, see Fig.\@ \ref{fig:fig_main}, allowing an analytical
evaluation of the observables. 

At small energy this hierarchy crosses over to hydrodynamic behaviour, see Fig.\@ \ref{fig:fig_main}, that we illustrate by calculating the local density of states. At
low energy it is suppressed in a power-law fashion according to the
Tomonaga-Luttinger theory. Away from the Fermi point where the Lorentz
invariance is reduced to Galilean by the parabolicity of the spectrum,
the local density of states is dominated by the first(leading)-level
excitations of the hierarchy. This produces a $1/\sqrt{\varepsilon}$
van Hove singularity, where $\varepsilon$ is the energy measured
from the bottom of the conduction band. At even higher energies the
second-level excitations produce another $1/\sqrt{\left|\varepsilon\right|}$
van Hove singularity on the other side of the band edge, in the forbidden
for the non-interacting system region.

Using this framework, we study response of the correlated system to
adding/removing a particle in detail, given by the spectral function.
The first-level excitations form a parabolic dispersion, like a single
particle, with a mass renormalised by the Luttinger parameter $K$.
\cite{TS14} The continuous spectrum of the second-level excitations
produces a power-law line-shape around the first-level mode with a
singular exponent $-1$. Around the spectral edges the second-level
excitations give a power-law behaviour of the spectral function. For
the hole edge the exponent calculated microscopically reproduces the
prediction of the phenomenological heavy impurity model in one-dimension.\cite{GlazmanReview12}
However, around the particle edge the second-level excitations give
a power-law of a new type.

Experimentally, momentum-resolved tunneling of electrons confined to a 1D geometry has been used to probe spin-charge separation in a Luttinger liquid.\cite{Yacoby02,Jompol09,Auslaender05,Tserkovnyak} This separation was observed to persist far beyond the energy range for which the Luttinger approximation is valid,\cite{Jompol09} showing the need for more sophisticated theories.\cite{Imambekov092} Particle-hole asymmetry has also been detected in relaxation processes.\cite{Yacoby10} In this paper we measure momentum-resolved tunneling of electrons in the upper layer of a GaAs-AlGaAs double-quantum-well structure from/to a 2D electron gas in the lower layer. This set-up probes the spectral function for spinful fermions. We observe well-resolved spin-charge separation at low energy with appreciable interaction strength---a distinct effect of the spinful generalisation of Luttinger liquid.\cite{GiamarchiBook} The ratio of charge and spin velocities is $v_{{\rm c}}/v_{{\rm s}}\approx1.8$.\cite{Jompol09} At high energy, in addition to the spin and charge curves, we can also resolve structure just above $k_{\rm F}$ that appears to be the edge of the second-level excitations. However, the amplitude decays rapidly and for higher $k$ we find no sign of the higher-level excitations, implying that their amplitude must have become at least three orders of magnitude weaker than for the parabola formed by the first-level excitations. The picture emerging out of these experimental results can only be explained---though only qualitatively---by the hierarchy that we study for spinless fermions.

The rest of the paper is organised as follows. In Section II we describe
the one-dimensional model of interacting spinless fermions introducing
a short range cut-off via lattice. Section III contains a procedure
of finding the many-body eigenenergy by means of the coordinate Bethe
ansatz. In Section IV we evaluate the form factors needed for the
dynamical response functions. We give a construction of the algebraic
representation of Bethe ansatz (Subsection IVa) and evaluate
the scalar product in this representation (Subsection IVb). We
present a calculation of the form factors for the spectral function
and the dynamical structure factor for a finite chain (Section IVc).
We take the limit of long wavelengths deriving polynomial formulae
for the form factors (Subsection IVd). Then, we analyse the obtained
form factors establishing hierarchy of excitations (Subsection IVe).
Finally we calculate the spectral function around the spectral
edges (Subsection IVf). In Section V we illustrate the crossover to Luttinger
liquid at low energy by evaluating the local density of states at
all energy scales. Section VI describes experiments on momentum-conserved
tunnelling of electrons in semiconductor wires. Section VII is dedicated
to low energies and in Section VIII we analyse the measurements at
high energies connecting the experiment with theory on spinless fermions
developed in this paper. Figures below are marked with spinless and spinful logos (such as those in Fig.\@ \ref{fig:SF_hierarchy_states} and \ref{fig:sc_separation}, respectively) to indicate the structure of the paper visually. Appendix A contains details of the derivation of the Bethe equations in the algebraic representation. In Appendix B we derive the expectation value of the local density operator.

\section{Model of spinless fermions}

We study theoretically the model of interacting Fermi particles without
spin in 1D, 
\begin{equation}
H=\int_{-\frac{L}{2}}^{^{\frac{L}{2}}}dx\left(-\frac{1}{2m}\psi^{\dagger}\left(x\right)\Delta\psi\left(x\right)+UL\rho\left(x\right)^{2}\right),\label{eq:H}
\end{equation}
where the field operators $\psi\left(x\right)$ satisfy the Fermi
commutation relations, $\left\{ \psi\left(x\right),\psi^{\dagger}\left(x'\right)\right\} =\delta\left(x-x'\right)$,
$\rho\left(x\right)=\psi^{\dagger}\left(x\right)\psi\left(x\right)$
is the particle density operator, and $m$ is the bare mass of a single
particle. Below we consider periodic boundary conditions, $\psi\left(x+L\right)=\psi\left(x\right)$,
restrict ourselves to repulsive interaction $U>0$ only, and take
$\hbar=1$.

Non-zero matrix elements of the interaction term in Eq.\@ (\ref{eq:H})
require a finite range of the potential profile for Fermi particles.
Here, we will introduce a lattice with next-neighbour interaction
which lattice parameter and interaction radius is $\mathcal{R}$.
The model in Eq.\@ (\ref{eq:H}) becomes 
\begin{equation}
H=\sum_{j=-\frac{\mathcal{L}}{2}}^{\frac{\mathcal{L}}{2}}\left[\frac{-1}{2m}\left(\psi_{j}^{\dagger}\psi_{j+1}+\psi_{j}^{\dagger}\psi_{j-1}\right)+U\rho_{j}\rho_{j+1}\right],\label{eq:H_lattice}
\end{equation}
where $j$ is the site index on the lattice, the dimensionless length
of the system is $\mathcal{L}=L/\mathcal{R}$, the operators obey
$\left\{ \psi_{j},\psi_{j}^{\dagger}\right\} =\delta_{ij}$, and $\rho_{j}=\psi_{j}^{\dagger}\psi_{j}$.

The long wavelength limit of the discrete model corresponds to the
model in Eq.\@ (\ref{eq:H}) while the interaction radius $\mathcal{R}$
provides microscopically an ultraviolet cutoff in the continuum regime.
For $N$-particle states of the lattice model we additionally impose
the constraint of low particle density, $N/\mathcal{L}\ll1$, to stay
within the conducting regime; a large occupancy $N\sim\mathcal{L}$
might lead to Wigner crystal physics at sufficiently strong interactions
that would localise the system. This procedure is analogous to the
point splitting regularisation technique \cite{vonDelftSchoellerReview}
which is usually introduced within the framework of the Luttinger
liquid mode in the linear regime.

\section{Spectral properties}

The model in Eq.\@ (\ref{eq:H_lattice}) can be diagonalised via
the Bethe ansatz approach which is based on the observation that the
eigenstates are superpositions of plain waves. This method is also
called coordinate Bethe ansatz.\cite{KorepinBook} The eigenstates,
following Ref.\@ \onlinecite{KorepinBook}, can be parameterised with
sets of $N$ quasimomenta $k_{j}$,
\begin{equation}
\Psi=\sum_{\mathcal{P},j_{1}<\dots<j_{N}}e^{i\sum_{l}k_{P_{l}}j_{l}+i\sum_{l<l'}\varphi_{P_{l},P_{l'}}}\psi_{j_{1}}^{\dagger}\dots\psi_{j_{N}}^{\dagger}\left|\textrm{vac}\right\rangle.\label{eq:psiN_coordinates}
\end{equation}
Their corresponding eigenenergies, $H\Psi=E\Psi$, are $E=\sum_{j=1}^{N}\left(1-\cos k_{j}\right)/m.$
Here $\left|\textrm{vac}\right\rangle $ is the vacuum state, the
scattering phases are fixed by the two-body scattering problem, 
\begin{equation}
e^{i2\varphi_{ll'}}=-\frac{e^{i\left(k_{l}+k_{l'}\right)}+1+2mUe^{ik_{l}}}{e^{i\left(k_{l}+k_{l'}\right)}+1+2mUe^{ik_{l'}}}\label{eq:phi_llp}
\end{equation}
and $\sum_{\mathcal{P}}$ is a sum over all permutation of quasimomenta.
The periodic boundary condition quantises the whole set of $N$ quasimomenta
simultaneously, 
\begin{equation}
\mathcal{L}k_{j}-2\sum_{l\neq j}\varphi_{jl}=2\pi I_{j}\label{eq:BA}
\end{equation}
where $I_{j}$ are sets of non-equal integer numbers. 

Generally, the system of equations in Eq.\@ (\ref{eq:BA}) has to be
solved numerically to obtain the full spectral structure of the observables.
However, in the long-wavelength regime the solutions can be evaluated
explicitly. 

In this limit the scattering phases in Eq.\@ (\ref{eq:phi_llp}) are
linear functions of quasimomenta, $2\varphi_{ll'}=\left(k_{l}-k_{l'}\right)/\left(1+\left(mU\right)^{-1}\right)+\pi$,
which makes the non-linear system of Bethe ansatz equations in Eq.
(\ref{eq:BA}) a linear system.\cite{TS14} Then, solving the linear
system for $\mathcal{L}\gg1$ via the matrix perturbation theory up
to the first subleading order in $1/\mathcal{L}$ we obtain
\begin{equation}
k_{j}=\frac{2\pi I_{j}}{\mathcal{L}-\frac{mUN}{mU+1}}-\frac{mU}{mU+1}\sum_{l\neq j}\frac{2\pi I_{l}}{\left(\mathcal{L}-\frac{mUN}{mU+1}\right)^{2}}.\label{eq:kj}
\end{equation}
Note that this calculation is valid for any interaction strength at
low densities. The corresponding eigenenergy and total momentum (protected
by the translational invariance of the system) are 
\begin{equation}
E=\sum_{j}\frac{k_{j}^{2}}{2m}\label{eq:E_kj}
\end{equation}
and $P=\sum_{j}k_{j}$. 

The spectrum of the many body states is governed by the first term
in Eq.\@ (\ref{eq:kj}). Reduction of the quantisation length in the
denominator of the first term in Eq.\@ (\ref{eq:kj}) is an exclusion
volume taken by the sum of interaction radii of all particles. Thus
all $N$-particle eigenstates at an arbitrary interaction strength
are given straightforwardly by the same sets of integer numbers $I_{j}$
as the free fermions' states, \emph{e.g.} the ground state corresponds
to $I_{j}=-N/2\dots N/2$.

For example, this result can be used to calculate the low energy excitations
explicitly that define the input parameters of the Luttinger-liquid
model, the velocity of the sound wave $v$ and of the Luttinger parameter
$K$. The first pair of the particle-like excitations, when an extra
electron is added just above the Fermi energy, have $I_{N+1}=N/2+1$
and $I_{N+1}=N/2+2$. The difference in their energies and momenta
are $E_{2}-E_{1}=\left(2\pi\right)^{2}N/\left[2m\left(\mathcal{L}-\frac{mUN}{mU+1}\right)^{2}\right]$
and $P_{2}-P_{1}=2\pi/\mathcal{L}$. Evaluating the discrete derivative,
which gives the slope of the dispersion around the Fermi energy, as
$v=\left(E_{2}-E_{2}\right)/\left(P_{2}-P_{1}\right)$ we obtain 
\begin{equation}
v=\frac{v_{\rm F}}{\left(1-\frac{NmU}{\mathcal{L}\left(1+mU\right)}\right)^{2}}\:\textrm{and}\:K=\left(1-\frac{NmU}{\mathcal{L}\left(1+mU\right)}\right)^{2},\label{eq:SF_Luttinger_params}
\end{equation}
where $v_{\rm F}=\pi N/\left(mL\right)$ is the Fermi velocity and the
relation $vK=v_{\rm F}$ between the Luttinger parameters for Galilean
invariant systems\cite{Haldane81} was used.

\section{matrix elements}

Now we turn to calculation of matrix elements. But first we need to
select operators that corresponds to specific observables. Our interest
lies in the dynamical response functions that correspond to adding/removing
a single particle to/from a correlated system and to creating an electron-hole
pair excitation out of the ground state of a correlated system. For
example, the first type of dynamics can be realised in experiments
using semiconductor nano-structures\cite{Yacoby02,Jompol09} where
an electrical current, generated by electrons tunnelling into/from
the nano-structure with their momentum and energy under controlled, probes
the system. 

The response of the many-body system to a single-particle excitation at
momentum $k$ and energy $\varepsilon$ is described by spectral function\cite{AGD}
$A\left(k,\varepsilon\right)=-\textrm{Im}\left[\int dxdte^{i\left(kx-\varepsilon t\right)}G\left(x,0,t\right)\right]\textrm{sgn}\left(\varepsilon-\mu\right)/\pi$.
Here $\mu$ is the chemical potential and $G\left(x,x',t\right)=-i\left\langle T\left(e^{-iHt}\psi\left(x\right)e^{iHt}\psi^{\dagger}\left(x'\right)\right)\right\rangle $
is Green function at zero temperature. In terms of the eigenstates
the spectral function reads 
\begin{multline}
A\left(k,\varepsilon\right)=L\sum_{f}\left|\left\langle f|\psi^{\dagger}\left(0\right)|0\right\rangle \right|^{2}\delta\left(\varepsilon-E_{f}+E_{0}\right)\delta\left(k-P_{f}\right)\\
+L\sum_{f}\left|\left\langle 0|\psi\left(0\right)|f\right\rangle \right|^{2}\delta\left(\varepsilon+E_{f}-E_{0}\right)\delta\left(k+P_{f}\right),\label{eq:A_continuum}
\end{multline}
where $E_{0}$ is the energy of the ground state $\left|0\right\rangle $,
and $P_{f}$ and $E_{f}$ are the momenta and the eigenenergies of
the eigenstates $\left|f\right\rangle $; all eigenstates are assumed
normalised. 

Creation of an electron-hole pair out of the correlated state at zero
temperature at momentum $k$ and energy $\varepsilon$ is described
by dynamical structure factor\cite{AGD} $S\left(k,\varepsilon\right)=\int dxdte^{i\left(kx-\varepsilon t\right)}\left\langle \rho\left(x,t\right)\rho\left(0,0\right)\right\rangle $,
where $\rho\left(x,t\right)=e^{-iHt}\rho\left(0\right)e^{iHt}$ is
the density operator evolving under the Hamiltonian to time $t$
and the average $\left\langle \dots\right\rangle $ is taken over
the ground state. In term of the eigenstates the dynamical structure
factor reads
\begin{equation}
S\left(k,\varepsilon\right)=L\sum_{f}\left|\left\langle f|\rho\left(0\right)|0\right\rangle \right|^{2}\delta\left(\varepsilon-E_{f}\right)\delta\left(k-P_{f}\right).\label{eq:S_continuum}
\end{equation}

Thus, we will be analysing the expectation values of the local operators
$\psi\left(0\right)$ and $\rho\left(0\right)$. To proceed with this
calculation we will borrow the result from Ref.\@ \onlinecite{Kitaine99, Kitaine00}
for Heisenberg chains. Our strategy is to perform the full calculation
for the discrete model in Eq.\@ (\ref{eq:H_lattice}) obtaining the
matrix elements of $\psi_{j}$ and $\rho_{j}$ as determinants. Then
we will take the long wavelength limit to evaluate the form factors
for the continuum model explicitly which will be the main technical
result in the theoretical part of this paper. Below we will construct
the algebraic form of Bethe ansatz, use the Slavnov's formula\cite{Slavnov89}
to express the scalar product and the normalisation factors in this
representation, and finally calculate the matrix elements of the local
operators.

\subsection{Algebraic Bethe ansatz}

The wave function of the $N$-particle eigenstates are factorised
in the algebraic representation which allows the general calculation
of various scalar products between them. Here we will follow the construction
in Ref.\@ \onlinecite{KorepinBook} for XXZ spins chains changing basis
from 1/2-spins to spinless fermions. 

The so called $R$-matrix acts on a tensor product $V_{1}\otimes V_{2}$
space and depends on an auxiliary parameter $u$, where $V_{1}$ and
$V_{2}$ are element-element subspaces each of which consists of two
states $\left|0\right\rangle _{j}$ and $\left|1\right\rangle _{j}$.
It is a solution of Yang-Baxter equation $R_{12}\left(u_{1}-u_{2}\right)R_{13}\left(u_{1}\right)R_{23}\left(u_{2}\right)=R_{23}\left(u_{2}\right)R_{13}\left(u_{1}\right)R_{12}\left(u_{1}-u_{2}\right)$.
For the lattice model in Eq.\@ (\ref{eq:H_lattice}) the $R$-matrix
reads 
\begin{multline}
R_{12}=1-\left(1-b\left(u\right)\right)\left(c_{1}^{\dagger}c_{1}+c_{2}^{\dagger}c_{2}\right)\\
-2b\left(u\right)c_{1}^{\dagger}c_{1}c_{2}^{\dagger}c_{2}+c\left(u\right)\left(c_{1}^{\dagger}c_{2}+c_{2}^{\dagger}c_{1}\right)
\end{multline}
where
\begin{equation}
b\left(u\right)=\frac{\sinh\left(u\right)}{\sinh\left(u+2\eta\right)},\:c\left(u\right)=\frac{\sinh\left(2\eta\right)}{\sinh\left(u+2\eta\right)}.\label{eq:bc_def}
\end{equation}
Here $\eta$ is the interaction parameter, and the tensor product space
is defined using fermionic basis $\left|0\right\rangle _{j}$ and
$\left|1\right\rangle _{j}$ with corresponding fermionic operators
$\left\{ c_{i},c_{j}^{\dagger}\right\} =\delta_{ij}$ that act in
these bases as $c_{j}^{\dagger}\left|0\right\rangle _{j}=\left|1\right\rangle _{j}$.
The latter will account for anticommuting nature of the lattice fermions
on different sites in contrast to the commutation relation of the
spin operators of a spin chain model.\cite{Korepin00, Sakai08} All
further calculation are identical to spin chains where the anti-commutation
relations of the Fermi particles are however automatically fulfilled.
This approach is more convenient than direct mapping of the results
for spin chains using Jordan-Wigner transformation.\cite{JordanWigner} 

A two-states subspace of $R$-matrix can be identified with the two-states
fermionic subspace of the lattice site $j$ of the model in Eq.\@ (\ref{eq:H_lattice}).
Then, the quantum version of the Lax operator (the so called $L$-matrix)
can be defined as $L_{j}=R_{\xi j}$. In the auxiliary subspace $\xi$
its matrix and operator forms are 
\begin{multline}
L_{j}=\left(\begin{array}{cc}
\frac{\cosh\left(u-\eta\left(2\rho_{j}-1\right)\right)}{\cosh\left(u-\eta\right)} & -i\frac{\sinh2\eta c_{j}^{-}}{\cosh\left(u-\eta\right)}\\
-i\frac{\sinh2\eta c_{j}^{\dagger}}{\cosh\left(u-\eta\right)} & -\frac{\cosh\left(u+\eta\left(2\rho_{j}-1\right)\right)}{\cosh\left(u-\eta\right)}
\end{array}\right),\\
=A^{j}\left(1-c_{\xi}^{\dagger}c_{\xi}\right)+c_{\xi}^{\dagger}C^{j}+B^{j}c_{\xi}+D^{j}c_{\xi}^{\dagger}c_{\xi}.\label{eq:L-matrix}
\end{multline}
Here the top left element element of the matrix is a transition between
$\left|0\right\rangle _{\xi}$ and $\left\langle 0\right|_{\xi}$
states of the auxiliary subspace, $c_{j}$ and $\rho_{j}$ are the
fermionic operators of the lattice model in Eq.\@ (\ref{eq:H_lattice}),
and $A^{j},B^{j},C^{j},D^{j}$ label the matrix elements of $L_{j}$.
The prefactor in front of $L_{j}$ was chosen such that for $u=i\pi/2-\eta$
it becomes a permutation matrix and for $\eta=0$ the $L$-operator
is diagonal.

By construction the $L$-operator satisfies algebra generated by Yang-Baxter
equation, 
\begin{equation}
R\left(u-v\right)\left(L_{j}\left(u\right)\otimes L_{j}\left(v\right)\right)=\left(L_{j}\left(v\right)\otimes L_{j}\left(u\right)\right)R\left(u-v\right).\label{eq:YB4L}
\end{equation}
The entries give commutation relations between the matrix elements
of $L$-matrix. Here we write down three of them that will be used
later, 
\begin{equation}
\left\{ B_{u}^{j},C_{v}^{j}\right\} =\frac{c\left(u-v\right)}{b\left(u-v\right)}\left(D_{v}^{j}A_{u}^{j}-D_{u}^{j}A_{v}^{j}\right),\label{BC}
\end{equation}
\begin{equation}
A_{u}^{j}C_{v}^{j}=\frac{1}{b\left(v-u\right)}C_{v}^{j}A_{u}^{j}-\frac{c\left(v-u\right)}{b\left(v-u\right)}C_{u}^{j}A_{v}^{j},\label{eq:AC}
\end{equation}
\begin{equation}
D_{u}^{j}C_{v}^{j}=-\frac{1}{b\left(u-v\right)}C_{v}^{j}D_{u}^{j}+\frac{c\left(u-v\right)}{b\left(u-v\right)}C_{u}^{j}D_{v}^{j}.\label{eq:DC}
\end{equation}
These relations can be also be checked explicitly by direct use of
the definition in Eq.\@ (\ref{eq:L-matrix}) and the Fermi commutation
relations.

The transition matrix $T\left(u\right)$ for a chain with $\mathcal{L}$
sites\textendash the so called monodromy matrix\textendash can
be defined similarly to the classical problem as 
\begin{equation}
T\left(u\right)=\sum_{j=1}^{\mathcal{L}}L_{j}\left(u\right).\label{eq:T-matrix}
\end{equation}
If all single-site $L$-matrices satisfy Eq.\@ (\ref{eq:YB4L}) then
the $T$-matrix also satisfies the same Yang-Baxter equation, e.g.
see proof in Ref.\@ \onlinecite{KorepinBook}. Therefore the matrix
elements of $T=A\left(1-c_{\xi}^{\dagger}c_{\xi}\right)+c_{\xi}^{\dagger}C+Bc_{\xi}+Dc_{\xi}^{\dagger}c_{\xi}$
in the 2x2 auxiliary space $\xi$ obey the same commutation relations
in Eqs. (\ref{BC}-\ref{eq:DC}). The transfer matrix for the whole chain,
\begin{equation}
\tau=\textrm{str}T=A\left(u\right)-D\left(u\right),\label{eq:transfer_matrix}
\end{equation}
is the super trace of $T$-matrix due to the fermionic definition
of the auxiliary space.\cite{Korepin00, Sakai08} The latter gives a
family of commuting matrices $\left[\tau\left(u\right),\tau\left(v\right)\right]=0$,
which contain all conserved quantities of the problem including the
Hamiltonian.

The vacuum state $\left|0\right\rangle$\textendash in the Fock
space of the model in Eq.\@ (\ref{eq:H_lattice})\textendash is an
eigenstate of the transfer matrix $\tau$. The corresponding eigenvalue,
$\tau\left(u\right)\left|0\right\rangle =\left(a\left(u\right)-d\left(u\right)\right)\left|0\right\rangle $,
is the difference of the eigenvalues of the $A$ and $D$ operators
which can be obtained directly by use of the definitions in Eqs.\@ (\ref{eq:L-matrix},
\ref{eq:T-matrix}). Noting that for $\mathcal{L}=2$ Eq.\@ (\ref{eq:T-matrix})
gives $A\left(u\right)\left|0\right\rangle =a_{1}\left(u\right)a_{2}\left(u\right)\left|0\right\rangle $
and $D\left(0\right)\left|0\right\rangle =d_{1}\left(u\right)d_{2}\left(u\right)\left|0\right\rangle $,
where $a_{1}\left(u\right)=a_{2}\left(u\right)=\cosh\left(u+\eta\right)/\cosh\left(u-\eta\right)$
and $d_{1}\left(u\right)=d_{2}\left(u\right)=1$, and generalising
this observation for arbitrary $\mathcal{L}$ one obtains
\begin{equation}
a\left(u\right)=\frac{\cosh\left(u+\eta\right)^{\mathcal{L}}}{\cosh\left(u-\eta\right)^{\mathcal{L}}},\quad\textrm{and}\quad d\left(u\right)=1.\label{eq:vacuum_ad}
\end{equation}

A general state of $N$ particles'\textendash Bethe state\textendash
is constructed by applying the operator $C\left(u\right)$ $N$ times
with different values of the auxiliary variable $u_{j}$, 
\begin{equation}
\Psi=\prod_{j=1}^{N}C\left(u_{j}\right)\left|0\right\rangle ,\label{eq:psiN_algebraic}
\end{equation}
where a set of $N$ values $u_{j}$ corresponds to $N$ quasimomenta
$k_{j}$ in coordinate representation of Bethe states in Eq.\@ (\ref{eq:psiN_coordinates}).
The state in Eq.\@ (\ref{eq:psiN_algebraic}) with an arbitrary set
of $u_{j}$ is not an eigenstate of the transfer matrix $\tau$. For
instance, it can be seen by commuting the operators $A$ and $D$
from left to right through all operators $C\left(u_{j}\right)$ that
generates many different states according to the commutation relations
in Eqs.\@ (\ref{eq:AC},\ref{eq:DC}). However the contribution of all of
the states that are non-degenerate with $\Psi$ can made to be zero by
choosing particular sets of $u_{j}$ that satisfy the following set
of equations 
\begin{equation}
\frac{a\left(u_{j}\right)}{d\left(u_{j}\right)}=\left(-1\right)^{N-1}\prod_{l=1\neq j}^{N}\frac{b\left(u_{l}-u_{j}\right)}{b\left(u_{j}-u_{l}\right)}\label{eq:BAequation_ABA}
\end{equation}
(see Appendix A for details). 

Under the substitution of the vacuum eigenvalues $a\left(u_{j}\right)$
and $d\left(u_{j}\right)$ of $A$ and $D$ operators from Eq.\@ (\ref{eq:vacuum_ad})
and $b\left(u_{l}-u_{j}\right)$ \textendash{} which define the commutation
relations Eqs. (\ref{eq:AC},\ref{eq:DC}) \textendash{} from Eq.
(\ref{eq:bc_def}) this so called eigenvalue equation above becomes
\begin{equation}
\frac{\cosh\left(u_{j}-\eta\right)^{\mathcal{L}}}{\cosh\left(u_{j}+\eta\right)^{\mathcal{L}}}=\left(-1\right)^{N-1}\prod_{l=1\neq j}^{N}\frac{\sinh\left(u_{j}-u_{l}-2\eta\right)}{\sinh\left(u_{j}-u_{l}+2\eta\right)}.\label{eq:eigenvalue_ABA}
\end{equation}
Thus all the sets of $u_{j}$ that satisfy the above equation give eigenstates
of the transfer matrix in the representation of Eq.\@ (\ref{eq:psiN_algebraic})
with the corresponding eigenvalues $\tau\left(u\right)\Psi=\mathcal{T}\left(u\right)\Psi$
where
\begin{equation}
\mathcal{T}\left(u\right)=a\left(u\right)\prod_{j=1}^{N}\frac{1}{b\left(u_{j}-u\right)}-\left(-1\right)^{N}d\left(u\right)\prod_{j=1}^{N}\frac{1}{b\left(u-u_{j}\right)}.\label{eq:tau_eigenvalue}
\end{equation}
This eigenvalue equation in the algebraic framework is the direct
analog of the Bethe ansatz equation (\ref{eq:BA}) in the coordinate
representation. Direct mapping between the two is done by the substitution
of 
\begin{equation}
e^{ik_{j}}=\frac{\cosh\left(u_{j}-\eta\right)}{\cosh\left(u_{j}+\eta\right)},\quad mU=-\cosh2\eta,\label{eq:ABA_to_CBA_mapping}
\end{equation}
in Eq.\@ (\ref{eq:BA}) and by taking its exponential.

!The original lattice Hamiltonian can be obtained from the
transfer matrix $\tau(u)$ that contains all of the conserved quantities of the problem.
Logarithmic derivatives of $\tau(u)$ give the global conservation laws by means of the so called trace identities, see  Ref.\@ \onlinecite{KorepinBook}. The linear coefficient in the Taylor series around the point $u=\frac{i\pi}{2}-\eta$ is proportional to the Hamiltonian itself. After restoring the correct prefactor the expression reads
\begin{equation}
H=-\frac{\sinh\eta}{2m}\partial_{u}\left.\ln\tau\left(u\right)\right|_{u=\frac{i\pi}{2}-\eta}.
\end{equation}
Substitution of the interaction parameter $\eta$ from Eq.\@ (\ref{eq:ABA_to_CBA_mapping}) in terms of the particle-particle interaction constant, $U$, in to the right hand side of the above relation recovers the lattice model in Eq.\@ (\ref{eq:H_lattice}).

\subsection{Scalar product}

The basic quantity, which calculations of expectation values will be based on, is the scalar product of two wave functions. A general way of evaluating it is the commutation relations in Eqs. (\ref{BC}-\ref{eq:DC}) and the vacuum expectation values of the $A$ and $D$ operators. The result of such a calculation simplifies greatly if one of the Bethe states is an eigenstate of the transfer matrix $\tau(u)$, as was first shown by Slavnov.\cite{Slavnov89} Then
the same result was rederived in Ref.\@ \onlinecite{Kitaine99, Kitaine00} using the so-called
factorising $F$-matrix,\cite{MailletSanchez00} which is a representation
of a Drinfeld twist.\cite{Drinfeld86} The latter will not be used in this Subsection but it will be needed later in  calculations of the matrix elements of the local operators.

Let $\left|\mathbf{u}\right\rangle =\prod_{j=1}^{N}C\left(u_{j}\right)\left|0\right\rangle $
be an eigenstate of the transfer matrix so that $N$ parameters
$u_{j}$ satisfy the Bethe equation in Eq.\@ (\ref{eq:eigenvalue_ABA}). And
let $\left\langle \mathbf{v}\right|=\left\langle 0\right|\prod_{j=1}^{N}B\left(v_{j}\right)$
be another Bethe state parametrised by a set of $N$ arbitrary values
$v_{j}$. The scalar product of these two states $\langle\mathbf{v}|\mathbf{u}\rangle$ can be evaluated
by commuting each operator $B\left(v_{j}\right)$ though the product
of $C\left(u_{j}\right)$ operators using the commutation relation
in Eq.\@ (\ref{BC}), which generates the $A$ and $D$ operators with all possible values of $u_j$ and $v_j$.
They, in turn, have also to be commuted to the right through the remaining
products of the $C\left(u_{j}\right)$ operators. Finally products of
the $A$ and $D$ operators, which act upon the vacuum state, just give
products of their vacuum eigenvalues $a(u_j), d(u_j)$ and $a(v_j), d(v_j)$ according to  Eq.\@ (\ref{eq:vacuum_ad}). 
The resulting sums of products can be written, using the relation between $u_j$ in Eq.\@ (\ref{eq:eigenvalue_ABA}), in a compact form as a determinant of an $N\times N$
matrix\cite{Slavnov89}
\begin{equation}
\left\langle \mathbf{v}|\mathbf{u}\right\rangle =\frac{\prod_{i,j=1}^{N}\sinh\left(v_{j}-u_{i}\right)}{\prod_{j<i}\sinh\left(v_{j}-v_{i}\right)\prod_{j<i}\sinh\left(u_{j}-u_{i}\right)}\det\hat{S}\label{eq:scalar_product}
\end{equation}
where the matrix elements are $S_{ab}=\partial_{u_{a}}\mathcal{T}\left(v_{b}\right)$.
Under substitution of the eigenvalues of the transfer matrix from
Eq.\@ (\ref{eq:tau_eigenvalue}) these matrix elements read in explicit
form as \begin{widetext} 
\begin{equation}
S_{ab}=-\frac{\cosh^{L}\left(v_{b}+\eta\right)}{\cosh^{L}\left(v_{b}-\eta\right)}\frac{\sinh\left(2\eta\right)}{\sinh^{2}\left(u_{a}-v_{b}\right)}\prod_{j=1\neq a}^{N}\frac{\sinh\left(u_{j}-v_{b}+2\eta\right)}{\sinh\left(u_{j}-v_{b}\right)}-\left(-1\right)^{N}\frac{\sinh\left(2\eta\right)}{\sinh^{2}\left(v_{b}-u_{a}\right)}\prod_{j=1\neq a}^{N}\frac{\sinh\left(v_{b}-u_{j}+2\eta\right)}{\sinh\left(v_{b}-u_{j}\right)}.\label{eq:scalar_product_matrix_elements}
\end{equation}
\end{widetext}

For $N=1$ the result in Eqs. (\ref{eq:scalar_product}, \ref{eq:scalar_product_matrix_elements})
follows directly from Eq.\@ (\ref{BC}). For arbitrary $N$ the proof
is more complicated: it employs the residue formula\cite{Slavnov89} (the function $\left\langle \mathbf{v}|\mathbf{u}\right\rangle$ has first order poles when $v_{i}\rightarrow u_{j}$) and the recurrent
relation for the scalar product of $N+1$ particles in terms of the
scalar product of $N$ particles, see also details
in Ref.\@ \onlinecite{KorepinBook}.

The normalisation factor of Bethe states can be obtained from Eq.
(\ref{eq:scalar_product}) by taking the limit $\mathbf{v}\rightarrow\mathbf{u}$.
The first order singularities, $\left(v_{b}-u_{b}\right)^{-1}$, in the off-diagonal matrix elements Eq.\@ (\ref{eq:scalar_product_matrix_elements})
are cancelled by zeros in the numerator in Eq.\@ (\ref{eq:scalar_product}).
The diagonal $a=b$ matrix elements contain second order singularities $\left(v_{b}-u_{b}\right)^{-2}$ for $\mathbf{v}\rightarrow\mathbf{u}$. However, the numerator also becomes zero when $\mathbf{v}\rightarrow\mathbf{u}$ in the leading order. Its expansion up to the first subleading order cancels the second order singularity of the denominator giving a finite expression for the matrix elements in the limit. The normalisation factor is found to be
\begin{equation}
\left\langle \mathbf{u}|\mathbf{u}\right\rangle =\sinh^{N}\left(2\eta\right)\prod_{i\neq j=1}^{N}\frac{\sinh\left(u_{j}-u_{i}+2\eta\right)}{\sinh\left(u_{j}-u_{i}\right)}\det\hat{Q} , \label{eq:norm}
\end{equation}
where the matrix elements are\begin{widetext} 
\begin{equation}
Q_{ab}=\begin{cases}
-\mathcal{L}\frac{\sinh2\eta}{\cosh\left(u_{a}+\eta\right)\cosh\left(u_{a}-\eta\right)}-\sum_{j\neq a}\frac{\sinh4\eta}{\sinh\left(u_{a}-u_{j}-2\eta\right)\sinh\left(u_{a}-u_{j}+2\eta\right)} & ,\; a=b,\\
\frac{\sinh4\eta}{\sinh\left(u_{b}-u_{a}+2\eta\right)\sinh\left(u_{b}-u_{a}-2\eta\right)} & ,\; a\neq b.
\end{cases}\label{eq:Qab}
\end{equation}
\end{widetext}

The last formula was originally derived by Gaudin using quantum mechanical
identities in the coordinate representation of Bethe ansatz.\cite{Gaudin1981}
Mapping of the resulting expression in Ref.\@ \onlinecite{Gaudin1981}
to the algebraic representation by means of Eq.\@ (\ref{eq:ABA_to_CBA_mapping})
gives directly the result in Eqs. (\ref{eq:scalar_product}, \ref{eq:scalar_product_matrix_elements})
with a different prefactors due to different normalisation factors
in the definitions of the states in Eq.\@ (\ref{eq:psiN_coordinates})
and of the states in Eq.\@ (\ref{eq:psiN_algebraic}). We will use the
algebraic form in Eq.\@ (\ref{eq:psiN_algebraic}) for the calculation
of the local matrix elements below.

\subsection{Expectation values of local operators}

Operators of the algebraic Bethe ansatz in Eqs. (\ref{eq:L-matrix}, \ref{eq:T-matrix})
are non-local in the basis of the original fermionic operators of the
lattice model in Eq.\@ (\ref{eq:H_lattice}). Thus, the first non-trivial
problem in calculating the matrix elements of the local operators $\psi_{j}^{\dagger}$ and $\rho_1$
in the algebraic representation of  Bethe states in Eq.\@ (\ref{eq:psiN_algebraic}) is
expressing the operators of our interest in terms of the non-local $A,B,C$ and $D$ operators from
Eqs. (\ref{eq:L-matrix}, \ref{eq:T-matrix}). Alternatively these Bethe operators
can be expressed in terms of the local operators of the lattice model.
The latter approach is much more complicated since the product of matrices
in Eq.\@ (\ref{eq:T-matrix}) is a large sum (exponential in the number of sites in the chain) 
restricting severely the ability to do explicit calculations
using the fermionic representation in practice.

An alternative way was found by constructing the $F$-matrix representation
of a Drinfeld twist.\cite{MailletSanchez00} In the $F$-basis the
monodromy matrix in Eq.\@ (\ref{eq:T-matrix}) becomes quasi-local, i.e.
its diagonal elements $A$ and $D$ become direct products of diagonal
matrices on each site over all sites of the chain and the off-diagonal
$B$ and $C$ are single sums over such direct products. Direct calculations
become much easier in this basis. Specifically, analysis of $A,B,C,D$
operators leads to a simple result for representing the $\psi_{j}$
operator in terms of algebraic Bethe ansatz operators, which then is shown to
be basis independent,\cite{Kitaine99, Kitaine00}
\begin{equation}
\psi_{j}^{\dagger}=\tau^{j-1}\left(\frac{i\pi}{2}-\eta\right)C\left(\frac{i\pi}{2}-\eta\right)\mathcal{\tau}^{\mathcal{L}-j}\left(\frac{i\pi}{2}-\eta\right).\label{eq:psij_aba}
\end{equation}
Here $\tau\left(u\right)=A\left(u\right)-D\left(u\right)$ is the super trace of the monodromy matrix and $C\left(u\right)$ is its matrix element.

The transfer matrices in the right hand side of the above equation
give only a phase prefactor in the expectation values with respect
to the Bethe states in Eq.\@ (\ref{eq:psiN_algebraic}). Let $\left|\mathbf{u}\right\rangle $
be an eigenstate of the transfer matrix with $N$ particles, let $\left|\mathbf{v}\right\rangle $
be an eigenstate with $N+1$ particles, and let us consider the case
of $j=1$. Acting with the $\tau^{\mathcal{L}-1}\left(i\pi/2-\eta\right)$
operator on the eigenstates $\left|\mathbf{u}\right\rangle $ gives
the eigenvalue $\prod_{j=1}^{N}\cosh^{\mathcal{L}-1}\left(u_{j}-\eta\right)/\cosh^{\mathcal{L}-1}\left(u_{j}+\eta\right)$
according to Eq.\@ (\ref{eq:tau_eigenvalue}). Then, using the mapping to the coordinate representation in Eq.\@ (\ref{eq:ABA_to_CBA_mapping}) and the Bethe equation in the form of Eq.\@ (\ref{eq:BA}), this eigenvalue can be expressed as  $\exp\left[i P_u \left(\mathcal{L}-1\right)\right]$ where $P_u$ is the total momentum of the state $u_j$, a quantum number. Similar phase factors for $j\neq 1$ are evaluated in an analogous way and each of them cancels out under modulus square in the form factor in Eq.\@ (\ref{eq:A_continuum}) making the local form factors independent of $j$ in full accord with the translational invariance of the system and the observable in Eq.(\ref{eq:A_continuum}). Thus
we will only calculate the value of $\left\langle \mathbf{v}|\psi_{1}^{\dagger}|\mathbf{u}\right\rangle $.  

Since $C\left(\frac{i\pi}{2}-\eta\right)\prod_{j=1}^{N}C\left(u_{j}\right)\left|0\right\rangle $
is also a Bethe state $\left|\frac{i\pi}{2}-\eta,u_{j}\right\rangle $,
though it is not an eigenstate, the expectation value can be calculated
using the result for the scalar product $\left\langle \mathbf{v}|\psi_{j}^{\dagger}|\mathbf{u}\right\rangle =\left\langle \mathbf{v}|\frac{i\pi}{2}-\eta,u_{j}\right\rangle $.
Substituting $\frac{i\pi}{2}-\eta,u_{j}$ in Eqs. (\ref{eq:scalar_product},\ref{eq:scalar_product_matrix_elements})
explicitly one obtains\begin{widetext} 
\begin{equation}
\left\langle \mathbf{v}|\psi_{1}^{\dagger}|\mathbf{u}\right\rangle =\left(-1\right)^{N+1}i\frac{\prod_{j=1}^{N+1}\cosh\left(v_{j}-\eta\right)}{\prod_{j=1}^{N}\cosh\left(u_{j}+\eta\right)}\frac{\sinh^{N+1}\left(2\eta\right)\det\hat{M}}{\prod_{j<i=2}^{N}\sinh\left(u_{j}-u_{i}\right)\prod_{j<i=2}^{N+1}\sinh\left(v_{j}-v_{i}\right)} \; ,\label{eq:psi_1_ABA}
\end{equation}
where the matrix elements are
\begin{eqnarray}
M_{ab} & = & \frac{\left(-1\right)^{N-1}}{\sinh\left(u_{b}-v_{a}\right)}\left(\prod_{j=1\neq b}^{N}\frac{\sinh\left(u_{b}-u_{j}+2\eta\right)}{\sinh\left(u_{b}-u_{j}-2\eta\right)}\prod_{j=1\neq a}^{N+1}\sinh\left(u_{b}-v_{j}-2\eta\right)+\prod_{j=1\neq a}^{N+1}\sinh\left(u_{b}-v_{j}+2\eta\right)\right) \; , \label{eq:Mab}
\end{eqnarray}
 \end{widetext}for $b<N+1$, and
\begin{eqnarray}
M_{ab} & = & \frac{1}{\cosh\left(v_{a}-\eta\right)\cosh\left(v_{a}+\eta\right)}\; , \label{eq:MaN1}
\end{eqnarray}
for $b=N+1$. Here the Bethe equation in Eq.\@ (\ref{eq:eigenvalue_ABA})
was used to express $a\left(v_{j}\right)/d\left(v_{j}\right)$ in
the matrix elements, and some factors in the matrix elements and the
overall prefactor cancel out. This result can be checked by numerical
evaluation of the sums over spacial variables using the coordinate
representation in Eq.\@ (\ref{eq:psiN_coordinates}) for a small number
of particles $N=1,2,3$ which we have done.

The determinant results in Eqs.\@ (\ref{eq:norm}, \ref{eq:scalar_product}, \ref{eq:psi_1_ABA})
can be checked by numerical evaluation of the sums over spacial variables
using the coordinate representation in Eq.\@ (\ref{eq:psiN_coordinates}).
However the latter summation over many coordinates has factorial complexity
which already limits numerical calculations to a few particles on
chains of a few dozens sites. The results of the algebraic Bethe
ansatz calculations have a power-law complexity that allows general
studies, at least numerically, of systems with hundreds of particles
on arbitrary long chains without making any approximations, e.g. the
studies of correlation functions in one-dimensional systems in Refs.
\onlinecite{JS05,JS05_SM, JS07,Pereira06,Pereira09,Links03}.

\subsection{The long wavelength limit}

We now turn to the evaluation of the long wavelength limit for matrix elements in the determinant form with the aim of calculating the determinants
explicitly. The resulting expressions will then be used to study physical
observables. 

Such an analysis is more convenient in the coordinate representation.
For small $k_{j}$ the non-linear mapping in Eq.\@ (\ref{eq:ABA_to_CBA_mapping})
becomes linear, similarly to Bethe equation in Eq.\@ (\ref{eq:BA})
in this limit. Then, a simple inversion of the linear function gives
\begin{equation}
u_{j}=\frac{i}{2}\sqrt{\frac{mU+1}{mU-1}}k_{j}\quad\textrm{and}\quad\eta=-\frac{1}{2}\textrm{acosh}\left(mU\right).\label{eq:ABA_CBA_lwl}
\end{equation}
Note that $\left|u_{j}\right|$ and $k_{j}$ are simultaneously much
smaller than one, while the interaction strength $U$ can be of an  arbitrary
magnitude.

We start from the expansion of the normalisation factor in Eq.\@ (\ref{eq:norm})
up to the leading non-vanishing order in the quasimomenta.
We first substitute Eq.\@ (\ref{eq:ABA_CBA_lwl}) in the matrix elements in
Eqs. (\ref{eq:Qab}), then expand them up the leading non-vanishing
order in $k_{j}\ll1$, and obtain  the diagonal matrix elements as follows, 
\begin{equation}
Q_{aa}=2\mathcal{L}\sqrt{\frac{mU-1}{mU+1}}-\frac{2\left(N-1\right)mU}{\sqrt{m^{2}U^{2}-1}} \; ,
\end{equation}
and 
\begin{equation}
Q_{ab}=\frac{2mU}{\sqrt{m^{2}U^{2}-1}} \; ,
\end{equation}
for $a\neq b$. The off-diagonal matrix elements are small compared
to the diagonal entries as $Q_{ab}/Q_{aa}\sim1/\mathcal{L}$ so the
leading contribution to the determinant is accumulated on the diagonal.
Also expanding the prefactor of Eq.\@ (\ref{eq:norm}) in small $k_{j}$
we obtain the following expression for the normalisation in the long
wavelength limit, 
\begin{equation}
\left\langle \mathbf{k}|\mathbf{k}\right\rangle =\frac{2^{N^{2}}\left(-1\right)^{N}\left(1-mU\right)^{N^{2}}\left(\mathcal{L}-\frac{mUN}{mU+1}\right)^{N}}{i^{N\left(N-1\right)}\prod_{i\neq j}\left(k_{j}-k_{i}\right)}\label{eq:norm_lwl}
\end{equation}
where $k_{j}$ are quasimomenta in the coordinate representation of
Bethe ansatz.

Our primary interest lies in the spectral function which contains
the local matrix element of $\psi^{\dagger}_j$ operators so here we
will focus on the determinant result in Eq.\@ (\ref{eq:psi_1_ABA}).
Similarly to the calculation of the normalisation factor we substitute
Eq.\@ (\ref{eq:ABA_CBA_lwl}) into Eq.\@ (\ref{eq:Mab}), which however
becomes zero in the zeroth order in $k_{j}$. Expanding it up to 
linear order in $k_{j}$ we obtain
\begin{equation}
M_{ab}=2mU\left(m^{2}U^{2}-1\right)^{\frac{N-1}{2}}\frac{\sum_{j=1}^{N}k_{j}^{u}-\sum_{j=1\neq a}^{N+1}k_{j}^{v}}{k_{b}^{u}-k_{a}^{v}}
\end{equation}
for $b<N+1$, where $\Delta P=\sum_{j}k_{j}^{u}-\sum_{j}k_{j}^{v}$
is the difference of two conserved quantities, the momenta of two states
$\mathbf{k}^{u}$ and $\mathbf{k}^{v}$. The matrix elements in Eq.
(\ref{eq:MaN1}) are already non-zero in the zeroth order in $k_{j}$
giving 
\begin{equation}
M_{ab}=\frac{2}{mU+1} \;,
\end{equation}
for $b=N+1$. Also expanding the prefactor in Eq.\@ (\ref{eq:psi_1_ABA})
and rearranging the expressions by taking a common factor out of the
matrix elements we obtain 
\begin{multline}
\left\langle \mathbf{k}^{v}|\psi^{\dagger}\left(0\right)|\mathbf{k}^{u}\right\rangle =\left(-1\right)^{N+1}i^{N^{2}}2^{N^{2}+N+\frac{1}{2}}\\1
\times\frac{\left(mU-1\right)^{N^{2}+\frac{1}{2}}m^{N}U^{N}\mathcal{D}}{\prod_{j<i}^{N}\left(k_{j}^{u}-k_{i}^{u}\right)\prod_{j<i}^{N+1}\left(k_{j}^{v}-k_{i}^{v}\right)}\label{eq:psi_1_lwl}
\end{multline}
where the entries of the matrix under the determinant, $\mathcal{D}=\det\hat{\mathcal{M}}$, for $b<N+1$ are 
\begin{equation}
\mathcal{M}_{ab}=\frac{\Delta P+k_{a}^{v}}{k_{b}^{u}-k_{a}^{v}}\quad\textrm{and}\quad\mathcal{M}_{a,N+1}=1.\label{eq:cMab}
\end{equation}

All matrix elements are of the the same order so the determinant in
Eq.\@ (\ref{eq:psi_1_lwl}) is a sum of a large number of terms unlike
the normalisation factor in Eq.\@ (\ref{eq:norm_lwl}). Doing the summation
we find an explicit expression in the form of a fraction of two polynomials in quasimomenta of the initial and the final states, 
\begin{multline}
\mathcal{D}=\left(-1\right)^{N+1}\frac{\prod_{j}\left(\Delta P+k_{j}^{u}\right)}{\prod_{i,j}\left(k_{j}^{v}-k_{i}^{u}\right)}\\
\prod_{j<i}^{N}\left(k_{j}^{u}-k_{i}^{u}\right)\prod_{j<i}^{N+1}\left(k_{j}^{v}-k_{i}^{v}\right).\label{eq:cD}
\end{multline}

For $N=1$ the result above is evaluated straightforwardly as a determinant
of a $2 \times 2$ matrix with the matrix elements in Eq.\@ (\ref{eq:cMab}). For
arbitrary $N$ we prove it by induction. Using the Laplace development
on the $N+1$ row, the determinant for $N+1$ particles can be expressed
as a sum of minors given, in turn, by determinants for $N$ particles,
$\mathcal{D}_{N+1}=\sum_{a=1}^{N+2}\left(-1\right)^{N+1+a}\mathcal{M}_{a,N+1}\textrm{minor}_{a,N+1}$,
which \textendash{} let us assume for purposes of the inductive method \textendash{} are given by Eq.\@ (\ref{eq:cD})
\begin{multline}
\textrm{minor}_{a,N+1}=\left(-1\right)^{N+1}\frac{\prod_{j=1}^{N}\left(\Delta P+k_{j}^{u}\right)}{\prod_{i=1,j=1\neq a}^{N,N+2}\left(k_{j}^{v}-k_{i}^{u}\right)}\\
\prod_{j<i}^{N}\left(k_{j}^{u}-k_{i}^{u}\right)\prod_{j<i\neq a}^{N+2}\left(k_{j}^{v}-k_{i}^{v}\right)
\end{multline}
Here $\mathcal{M}_{a,N+1}$ are given by the  matrix elements in Eq.
(\ref{eq:cMab}), $N$ quasimomenta $k_{j}^{u}$ are labeled by $j=1\dots N$,
and $N+1$ quasimomenta $k_{j}^{v}$ are labeled by $j=1\dots N+2$
with $a^{th}$ elements excluded. Under taking common factor in front
of the sum, the determinant for $N+1$ particles becomes
\begin{widetext}
\begin{multline}
\mathcal{D}_{N+1}=\left(-1\right)^{N+2}\frac{\prod_{j=1}^{N+1}\left(\Delta P+k_{j}^{u}\right)\prod_{j<i}^{N+1}\left(k_{j}^{u}-k_{i}^{u}\right)\prod_{j<i}^{N+2}\left(k_{j}^{v}-k_{i}^{v}\right)}{\prod_{i,j}\left(k_{j}^{v}-k_{i}^{u}\right)}\\
\times\frac{1}{\Delta P+k_{N+1}^{u}}\sum_{a=1}^{N+2}\frac{\left(\Delta P+k_{a}^{v}\right)\prod_{j=1\neq a}^{N+2}\left(k_{j}^{v}-k_{N+1}^{u}\right)\prod_{j=1}^{N}\left(k_{j}^{u}-k_{a}^{v}\right)}{\prod_{j=1}^{N}\left(k_{j}^{u}-k_{N+1}^{u}\right)\prod_{j=1\neq a}^{N+2}\left(k_{j}^{v}-k_{a}^{v}\right)}.\label{eq:cDN1}
\end{multline}
\end{widetext}The sum in the above expression gives, by direct calculation,
$\sum_{a=1}^{N+2}\dots=\Delta P+k_{N+1}^{u}$ which makes the whole
second line unity. The determinant is equal to the first line
of Eq.\@ (\ref{eq:cDN1}) which is also equal to the result in Eq.\@ (\ref{eq:cD})
for $N+1$ particles. Thus we obtained the same result for $N+1$
particles starting from Eq.\@ (\ref{eq:cD}) for $N$ particles. Hence, it is proved by
induction.

Finally, the form factor in the Eq.\@ (\ref{eq:A_continuum}) is the
modulus squared of Eq.\@ (\ref{eq:psi_1_lwl}). Normalising the initial
and the final state wave functions using Eq.\@ (\ref{eq:norm_lwl})
as $\left|\left\langle f|\psi^{\dagger}\left(0\right)|0\right\rangle \right|^{2}=\left|\left\langle \mathbf{k}^{f}|\psi^{\dagger}\left(0\right)|\mathbf{k}^{0}\right\rangle \right|^{2}\left\langle \mathbf{k}^{f}|\mathbf{k}^{f}\right\rangle ^{-1}\left\langle \mathbf{k}^{0}|\mathbf{k}^{0}\right\rangle ^{-1}$
we obtain

\begin{multline}
\left|\left\langle f|\psi^{\dagger}\left(0\right)|0\right\rangle \right|^{2}=\frac{Z^{2N}}{\mathcal{L}}\frac{\prod_{j}^{N}\left(k_{j}^{0}-P_{f}\right)^{2}}{\prod_{i,j}^{N,N+1}\left(k_{j}^{f}-k_{i}^{0}\right)^{2}}\\
\prod_{i<j}^{N}\left(k_{j}^{0}-k_{i}^{0}\right)^{2}\prod_{i<j}^{N+1}\left(k_{j}^{f}-k_{i}^{f}\right)^{2},\label{eq:FF_N}
\end{multline}
where $Z=mU/\left(mU+1\right)/\left(\mathcal{L}-NmU/\left(1+mU\right)\right)$,
$k_{j}^{f}$ and $k_{j}^{0}$ are the quasimomenta of the eigenstate
$\left|f\right\rangle $ and the ground state $\left|0\right\rangle $,
and $P_{0}=0$ for the ground state.

Calculation of $\left\langle f|\rho\left(0\right)|0\right\rangle $
is done in a similar way by expressing the local density operator
$\rho_{1}$, within the framework of the lattice model, in terms of
the algebraic Bethe ansatz operators $A$, $B$, $C$, $D$ and, then,
by using the Slavnov formula. Details are given in appendix B. In
the long wavelength limit we obtain
\begin{multline}
\left|\left\langle f|\rho\left(0\right)|0\right\rangle \right|^{2}=\frac{Z^{2N-2}}{\mathcal{L}^{2}}\frac{P_{f}^{2N}}{\prod_{i,j}^{N,N}\left(k_{j}^{f}-k_{i}^{0}\right)^{2}}\\
\prod_{i<j}^{N}\left(k_{j}^{0}-k_{i}^{0}\right)^{2}\prod_{i<j}^{N}\left(k_{j}^{f}-k_{i}^{f}\right)^{2},\label{eq:FF_dsf}
\end{multline}
where the final states $\left|f\right\rangle $ have the same number
of excitations $N$ as the ground state $\left|0\right\rangle $,
unlike in Eq.\@ (\ref{eq:FF_N}), and $P_{0}=0$ for the ground state
as in Eq.\@ (\ref{eq:FF_N}).

These form factors in Eqs. (\ref{eq:FF_N}, \ref{eq:FF_dsf}) together
with the solution of Bethe equations in Eq.\@ (\ref{eq:kj}) is the
main technical result in the theory part of our work. We will analyse
its physical consequences in the next two Subsections. The similarity between these
two expressions means that the hierarchy of modes we will identify below is a general feature
of one and two body operators.

\subsection{Hierarchy of modes}

The results in Eqs. (\ref{eq:FF_N}, \ref{eq:FF_dsf}) have one or
more singularities when one or more quasimomenta of an excited state
coincide with a quasimomentum of the ground state, $k_{j}^{f}=k_{j}^{0}$. Both results have a multiplicand $Z^{2N}\sim\mathcal{L}^{-2N}$ that becomes
virtually zero in the thermodynamic limit, in which $\mathcal{L}\rightarrow\infty$.
Thus the product of these two opposite factors  produces an uncertainty in the
limiting behaviour (of the $0\times\infty$ type) that has to be resolved.
Since we are specifically interested in a transport experiment in
this paper, in which the spectral function is measured, we will mainly focus
on solving the uncertainty problem for result in Eq.\@ (\ref{eq:FF_N}).
\begin{figure}
{\centering\includegraphics[width=0.95\columnwidth]{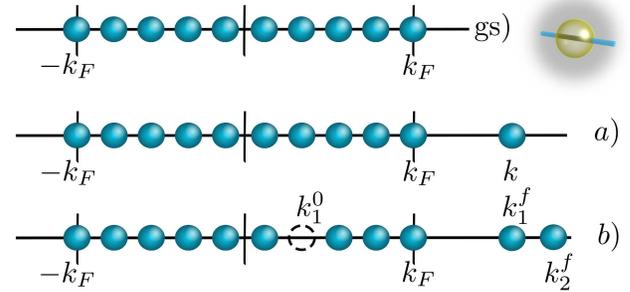}}
\caption{\label{fig:SF_hierarchy_states} Configurations of quasimomenta that solve the Bethe equations in Eqs. (\ref{eq:BA}, \ref{eq:phi_llp})  for the spinless fermions model in Eq.\@ (\ref{eq:H_lattice}):
gs) the ground state, a) excitations that form the $a$-level of the hierarchy, and b)
excitations that form the $b$-level of the hierarchy.}
\end{figure}

The maximum number of singularities is $N$ in the extreme case,  when quasimomenta
$k_{j}^{f}$ of an excited state coincide with all of the $N$ quasimomenta of the
ground state $k_{j}^{0}$ given in Fig.\@ \ref{fig:SF_hierarchy_states}gs.
The excited states of this kind are given in Fig.\@ \ref{fig:SF_hierarchy_states}a.
The divergences in the denominator of Eq.\@ (\ref{eq:FF_N}) occur only in the leading order---the first term in Eq.\@ (\ref{eq:kj})---but the subleading order---the second term in Eq.\@ (\ref{eq:kj})---already provides a self-consistent
cutoff within the theory. The interaction shift of the quasimomenta at subleading order does not cancel for the extra added particle in
the excited state, making the factors in the denominator of Eq.\@ (\ref{eq:FF_N})
\begin{equation}
k_{j}^{f}-k_{j}^{0}=\frac{mU}{mU+1}\frac{k_{N+1}^{f}-k_{j}^{0}}{\mathcal{L}-\frac{mUN}{mU+1}},\label{eq:kfk0_denom}
\end{equation}
where in the r.h.s. only the first  term from Eq.\@ (\ref{eq:kj})
is relevant for $k^f_{N+1}$ and $k^0_j$. The numerator for the states in Fig.\@ \ref{fig:SF_hierarchy_states}a
becomes 
\begin{equation}
k_{j}^{0}-P_{f}=k_{j}^{0}-k_{N+1}^{f}.\label{eq:k0Pf_num}
\end{equation}
Substitution of Eqs. (\ref{eq:kfk0_denom}, \ref{eq:k0Pf_num}) in
Eq.\@ (\ref{eq:FF_N}) for one particle, say for $j=N$, cancels one factor
$Z^2\sim\mathcal{L}^{-2}$ and the other part of the product for $i\neq j$
in the denominator of first line of Eq.\@ (\ref{eq:FF_N}) cancels partially
the products in the second line of Eq.\@ (\ref{eq:FF_N}). The expression
for the remaining $N-1$ particles is the same as Eq.\@ (\ref{eq:FF_N})
but the numbers of terms in the products are reduced by one, $N\rightarrow N-1$,
giving 
\begin{multline}
\left|\left\langle f|\psi^{\dagger}\left(0\right)|0\right\rangle \right|^{2}=\frac{Z^{2N-2}}{\mathcal{L}}\frac{\prod_{j}^{N-1}\left(k_{j}^{0}-P_{f}\right)^{2}}{\prod_{i,j}^{N-1,N}\left(k_{j}^{f}-k_{i}^{0}\right)^{2}}\\
\prod_{i<j}^{N-1}\left(k_{j}^{0}-k_{i}^{0}\right)^{2}\prod_{i<j}^{N}\left(k_{j}^{f}-k_{i}^{f}\right)^{2}.\label{eq:FF_Nm1}
\end{multline}
Repeating the procedure $N-1$ times we cancel the remaining $Z^{2N-2}$
factor completely (with the rest of other terms) and obtain 
\begin{equation}
\mathcal{L}\left|\left\langle f|\psi^{\dagger}\left(0\right)|0\right\rangle \right|^{2}=1.\label{eq:Aa}
\end{equation}
Corrections to this result originate from higher subleading orders  in the
solutions to Bethe equations in Eq.\@ (\ref{eq:kj}) and are of the
order of $O\left(\mathcal{L}^{-1}\right)$. This becomes much smaller
than one at leading order of Eq.\@ (\ref{eq:Aa}) in the thermodynamic limit.

Substitution of Eq.\@ (\ref{eq:Aa}) in Eq.\@ (\ref{eq:A_continuum})
gives the value of the spectral function $A\left(k,\varepsilon\left(k\right)\right)=1$.
The energies and the momenta of the excitations in Fig.\@ \ref{fig:SF_hierarchy_states}a form a single line on the spectral plane, like a single particle with dispersion
$\varepsilon\left(k\right)=k^{2}/\left(2m^{*}\right)$, where the
effective mass is renormalised by the Luttinger parameter $K$, $m^{*}=mK$.\cite{TS13}
Note that, since we still resolve individual levels here, the delta functions in the definition of the spectral function in Eq.\@ (\ref{eq:A_continuum}) become discrete Kronecker deltas.
Thus,  $A\left(k,\varepsilon\right)$
at each discrete point $k,\varepsilon$ describes the probability of adding
(removing) a particle, which is non-negative and is bound by one from above, instead of the probability density as in the continuum case. Dimensional analysis makes this distinction clear immediately.

The excitations that have one singularity less ($N-1$ in total) can
be  visualised systematically as an extra electron-hole pair created
in addition to adding an extra particle, see Fig.\ref{fig:SF_hierarchy_states}b. Staring from
Eq.\@ (\ref{eq:FF_N}) and using the same procedure as before Eq.\@ (\ref{eq:FF_Nm1}) but
$N-1$ instead of $N$ times we obtain 
\begin{equation}
\left|\left\langle f|\psi^{\dagger}\left(0\right)|0\right\rangle \right|^{2}=\frac{Z^{2}}{\mathcal{L}}\frac{\left(k_{2}^{f}-k_{1}^{f}\right)^{2}\left(k_{1}^{0}-P_{f}\right)^{2}}{\left(k_{1}^{f}-k_{1}^{0}\right)^{2}\left(k_{2}^{f}-k_{1}^{0}\right)^{2}},\label{eq:FF_2}
\end{equation}
where $k_{1}^{f},k_{2}^{f}$ and $k_{1}^{0}$ are positions of two
particles and one hole in Fig.\@ \ref{fig:SF_hierarchy_states}b. Substitution of Eq.\@ (\ref{eq:FF_2})
in Eq.\@ (\ref{eq:A_continuum}) gives values of the spectral function
$A\left(k,\varepsilon\right)\sim\mathcal{L}^{-2}$ that 
smaller than the values for the excitations in Fig.\@ \ref{fig:SF_hierarchy_states}a  (in Eq.\@ (\ref{eq:A_continuum})) by a factor of $\mathcal{L}^{-2}$.
For two singularities less ($N-2$ in total) we find $A\left(k,\varepsilon\right)\sim\mathcal{L}^{-4}$
and so on.

\begin{figure}
\begin{centering}
\includegraphics[width=0.95\columnwidth]{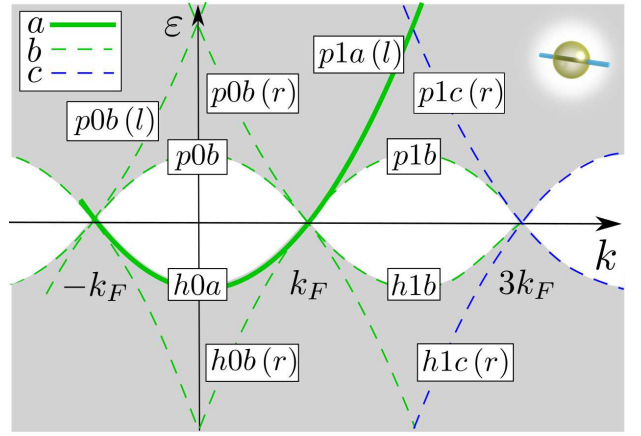} 
\end{centering}
\protect\caption{
\label{fig:SF_spectral_function}The spectral function for interacting spinless fermions in the region
$-k_{\rm F}<k<k_{\rm F}$ ($k_{\rm F}<k<3k_{\rm F}$) labelled
by $0\left(1\right)$. The grey areas mark non-zero values. The green and the blue lines are modes of the hierarchy labelled as follows: $p\left(h\right)$
shows the particle(hole) sector, $k_{\rm F}$ is the Fermi momentum,
$a,\;b,\;c$ respectively identify the level in the hierarchy in powers
$0,\;1,\;2$ of $\mathcal{R}^{2}/L^{2}$, and $\left(r,l\right)$
specifies the origin in the range\textemdash modes on the edge have
no such label.}
\end{figure}

This emerging structure separates the plethora of many-body excitations into  a hierarchy according to the remaining powers of $\mathcal{L}^{-2}$ in their respective form factors. We label the levels of the hierarchy as $a,\;b,\;c$ reflecting
the factors $\mathcal{L}^{-2n}$ with $n=0,\;1,\;2$. While the leading $a$-excitations form a discrete single-particle-like dispersion, see $h0a$ and $p1a(l)$ lines in Fig.\@ \ref{fig:SF_spectral_function}, the spectral properties of the  subleading excitations are described by a more complicated continuum of states  on the energy-momentum plane. We will explore the $b$-excitations below. 

All simple modes, formed by single particle- and hole-like excitations  of the type
in Fig.\@ \ref{fig:SF_hierarchy_states}b, in the range $-k_F<k<3 k_F$ are presented in Fig.\@ \ref{fig:SF_spectral_function}. We use the following naming scheme:   $p\left(h\right)$ indicates the particle
(hole) sector, $0\left(1\right)$ encodes the range of momenta $-k_{\rm F}<k<k_{\rm F}$
$\left(k_{\rm F}<k<3k_{\rm F}\right)$,  $a,\;b,\;c$ reflect
the terms $\mathcal{L}^{-2n}$ with $n=0,\;1,\;2$. 
The suffix  $\left(r\right)$ or $\left(r\right)$ marks a particle-like mode, e.g. the states with in Fig.\@ \ref{fig:SF_hierarchy_states}b with $k_{1}^{f}=-k_{\rm F}-\gamma$, $k_{1}^{0}=k_{\rm F}$,
and $k_{\rm F}>P_{f}=-2k_{\rm F}+k_{2}^{f}>-k_{\rm F}$ forms the mode $p0b(l)$. Hole-like modes have no suffixes, e.g.  the states  in Fig.\@ \ref{fig:SF_hierarchy_states}b with $k_{1}^{f}=-k_{\rm F}-\gamma$,
$k_{2}^{f}=k_{\rm F}+\gamma$, and $-k_{\rm F}<P_{f}=-k_{1}^{0}<k_{\rm F}$ form the mode $p0b$.  Simple modes formed by excitations of lower levels of the hierarchy are obtained by translation of the $b$-modes constructed in this paragraph by integer numbers of $\pm 2 k_F$. 
A couple of simple modes formed by $c$-excitations are  presented on Fig.\@ \ref{fig:SF_spectral_function}. They have the same naming scheme  as  the $b$-modes.

Now we evaluate values of  the spectral function along all simple $b$-modes in the
range $-k_{\rm F}<k<3k_{\rm F}$. Let us start from the $p0b$ mode, see Fig.\@
\ref{fig:SF_spectral_function}. Along this mode the spectral function
is a bijective function of $k$, $A\left(k,\varepsilon_{p0b}\left(k\right)\right)$
where $\varepsilon_{p0b}\left(k\right)=k_{\rm F}^{2}/\left(mK\right)-k^{2}/\left(2mK\right)$.
The states that form it belong to $b$-excitations in Fig.\@ \ref{fig:SF_hierarchy_states}b
with $k_{1}^{f}=-k_{\rm F}-\gamma$, $k_{2}^{f}=k_{\rm F}+\gamma$, and $k=P_{f}=-k_{1}^{0}$.
Substituting this parameterisation in Eq.\@ (\ref{eq:FF_2}) we obtain
\begin{equation}
A\left(k,\varepsilon_{p0b}\left(k\right)\right)=\frac{16Z^{2}k_{\rm F}^{2}k^{2}}{\left(k^{2}-\left(k_{\rm F}+\gamma\right)^{2}\right)^{2}}.\label{eq:Ap0b}
\end{equation}
The spectral function along all other $b$-modes in Fig.\@ \ref{fig:SF_spectral_function}
is calculated in the very same way and the results (together with
$a$-modes) are summarised in Table \ref{tab:SF_spectral_function_values}.
\begin{table}
\begin{ruledtabular}
\begin{tabular}{r|c|c}
\multicolumn{1}{r|}{$\textrm{ }$ } & $x=0$  & $x=1$\tabularnewline
\hline 
$pxa$  & $-$  & $1$\tabularnewline
\hline 
$hxa$  & $1$  & $-$\tabularnewline
\hline 
$pxb$  & $\frac{16Z^{2}k_{\rm F}^{2}k^{2}}{\left(k^{2}-\left(k_{\rm F}+\gamma\right)^{2}\right)^{2}}$  & $\frac{4Z^{2}\gamma^{2}\left(k-k_{\rm F}+\frac{3}{2}\gamma\right)^{2}}{\left(k-k_{\rm F}+\gamma\right)^{2}\left(k-k_{\rm F}+2\gamma\right)^{2}}$\tabularnewline
\hline 
$pxb\left(l\right)$  & $\frac{4Z^{2}\left(k_{\rm F}+k\right)^{2}}{k_{\rm F}^{2}}$  & $-$\tabularnewline
\hline 
$pxb\left(r\right)$  & $\frac{4Z^{2}\left(k_{\rm F}-k\right)^{2}}{k_{\rm F}^{2}}$  & $-$\tabularnewline
\hline 
$hxb$  & $-$  & $\frac{4Z^{2}\left(3k_{\rm F}-k-\gamma\right)^{2}\left(k_{\rm F}+k\right)^{2}}{k_{\rm F}^{2}\left(k-k_{\rm F}+\gamma\right)^{2}}$\tabularnewline
\hline 
$hxb\left(l\right)$  & $\frac{4Z^{2}\gamma^{2}}{\left(k+k_{\rm F}+2\gamma\right)^{2}}$  & $\frac{Z^{2}k_{\rm F}^{2}k^{2}}{\left(\left(k+\gamma\right)^{2}-k_{\rm F}^{2}\right)^{2}}$\tabularnewline
\hline 
$hxb\left(r\right)$  & $\frac{4Z^{2}\gamma^{2}}{\left(k-k_{\rm F}-2\gamma\right)^{2}}$  & $-$\tabularnewline
\end{tabular}
\end{ruledtabular}
\caption{\label{tab:SF_spectral_function_values}Spectral weights along the
$a$- and the $b$-modes for $-k_{\rm F}<k<k_{\rm F}$ ($k_{\rm F}<k<3k_{\rm F}$)
labeled by $x=0\left(1\right)$. Terminology is the same as in Fig.\@
\ref{fig:SF_spectral_function}; $\gamma=2\pi/\mathcal{L}$ and  $Z=mU/\left(mU+1\right)/\left(\mathcal{L}-NmU/\left(1+mU\right)\right)$.}
\end{table}

The amplitude of the subleading $b$-excitations
does not vanish in the thermodynamic limit, though it is proportional
to $1/\mathcal{L}$. The limit involves both ${\cal L\rightarrow\infty}$
and the particle number $N\rightarrow\infty$ but keeps the density
$N/{\cal L}$ finite. The spectral weights of the subleading modes
$p0b$, $h1b$, and $h1b\left(r\right)$ from Table \ref{tab:SF_spectral_function_values}
are proportional to the density squared for some values for $k$,
e.g. the modes $p0b$ at $k=k_{\rm F}$ gives 
\begin{equation}
A\left(k_{\rm F},\varepsilon_{p0b}\left(k_{\rm F}\right)\right)=\left(\frac{mU}{1+mU}\right)^{2}\frac{N^{2}}{\left(\mathcal{L}-\frac{NmU}{1+mU}\right)^{2}},
\end{equation}
see Table \ref{tab:SF_spectral_function_values} for other modes, and are apparent in the infinite system.

Assessing further the continuum of $b$-excitations we consider the spectral function
and how it evolves as one moves away slightly from the 
of the strongest $a$-mode. Just a single step of a single quantum
of energy away from the $a$-mode requires the addition of an electron-hole
pair on top of the configuration of quasimomenta in Fig.\@ \ref{fig:SF_hierarchy_states}a.
This immediately moves such states one step down the hierarchy to
$b$-excitations. Let us consider the spectral function as a function
of energy only making a cut along a line of constant $k$. The energies
of the electron-hole pairs themselves are regularly spaced around
the Fermi energy with slope $v_{\rm F}$. However, the degeneracy of
the many-body excitations due to the spectral linearity makes the
level spacings non-equidistant. We smooth this irregularity using
an averaging of the spectral function over energy, 
\begin{equation}
\overline{A}\left(k,\varepsilon\right)=\int_{-\frac{\epsilon_{0}}{2}}^{\frac{\epsilon_{0}}{2}}\frac{d\epsilon}{\epsilon_{0}}A\left(k,\varepsilon+\epsilon\right)\label{eq:average}
\end{equation}
where $\epsilon_{0}$ is a small energy scale. 

Then, using the parametrisation
of $b$-excitations in Fig.\@ \ref{fig:SF_hierarchy_states}b in the
vicinity of the principal parabola, we linearise the energies of the
extra electron-hole pairs around the Fermi energy and of the particle
around its original position. We then substitute the resulting expressions
for $k_{1,2}^{f}$ and $k_{1}^{0}$ in terms of the energy $E$ from
Eq.\@ (\ref{eq:E_kj}) in Eq.\@ (\ref{eq:FF_2}), similar to our procedure
of obtaining Eq.\@ (\ref{eq:Ap0b}). Finally, we use the averaging rule
in Eq.\@ (\ref{eq:average}) and obtain 
\begin{eqnarray}
\overline{A}\left(k,\varepsilon\right) & = & \frac{Z^{2}2k_{\rm F}\left(3k^{2}+k_{\rm F}^{2}\right)\theta\left(\varepsilon_{h0a}\left(k\right)-\varepsilon\right)}{m\gamma K\left(\varepsilon_{h0a}\left(k\right)-\varepsilon\right)}\label{eq:Ashape0}\\
 &  & \qquad\qquad\qquad\qquad\textrm{for}\;-k_{\rm F}<k<k_{\rm F},\nonumber \\
\overline{A}\left(k,\varepsilon\right) & = & \frac{Z^{2}\left(k+\textrm{sgn}\left(\varepsilon-\varepsilon_{p1a\left(l\right)}\left(k\right)\right)k_{\rm F}\right)^{3}}{m\gamma K\left|\varepsilon-\varepsilon_{p1a\left(l\right)}\left(k\right)\right|}\label{eq:Ashape1}\\
 &  & \qquad\qquad\qquad\qquad\textrm{for}\;k_{\rm F}<k<3k_{\rm F}\nonumber 
\end{eqnarray}
where $\gamma=2\pi/\mathcal{L}$ and $\varepsilon_{h0a}\left(k\right)=\varepsilon_{p1a\left(l\right)}\left(k\right)=k^{2}/\left(2mK\right)$
is the parabolic dispersion of the $a$-mode.

The result in Eqs. (\ref{eq:Ashape0}, \ref{eq:Ashape1}) can be interpreted
as the line-shape of the $a$-mode. However, it has an unusual
form---namely that of a divergent power-law. The divergence at the parabola is cut-off 
by the lattice spacing recovering $\overline{A}\left(k,\varepsilon_{h0a,p1a}\left(k\right)\right)=1$
from Eq.\@ (\ref{eq:Aa}). In Eq.\@ (\ref{eq:Ashape1}) the line-shape
is asymmetric due to different prefactors $\left(k\pm k_{\rm F}\right)^{3}$
above and below the line. In Eq.\@ (\ref{eq:Ashape0}) the higher energy
part ($\varepsilon>\varepsilon_{h0a}\left(k\right)$) is absent due
to the absence of the excitation in this region, forbidden by the kinematic
constraint.

Not every simple mode marks a distinct feature. The states at
least on one side, above or below the mode in energy, have to belong
to a different level of the hierarchy than the mode itself, which
results in a divergence or in a jump of the spectral function in the
continuum of excitations. Otherwise the spectral function is continuous
across all of the modes that belong to the same level of the hierarchy
as the excitations around them. The $a$-modes are distinct since excitations
around them belong to a different $b$-level. All  modes on the spectral edges,  $p0b$,
$p1b$,  $h1b$, and so on, are distinct since on one side there are no excitations
(due to the kinematic constraint) and on the other side there is
a finite density of states resulting in a jump of the spectral function. 

An example of an observable subleading mode in the continuum is $h0b\left(r\right)$. 
On the higher energy side of this mode,  the excitations are
described by the same type of states in Fig.\@ \ref{fig:SF_hierarchy_states}b
but on the lower energy side creation of an additional electron-hole
pair in the quasimomenta results in states that have two non-cancelled
singularities in Eq.\@ (\ref{eq:FF_N}), which lowers their corresponding
level of the hierarchy to $c$ from $b$. This, in turn, results  in an observable
feature in the spectral function at the position of the $h0b\left(r\right)$ mode.
On the other hand, the $p0b\left(r\right)$ and $h1b\left(l\right)$ modes in  continuum are not detectable since excitations on both sides around them belong
to the same $b$-level of the hierarchy. Observability of all other
modes can be easily assessed in the same way by considering their
corresponding states in the form of Fig.\@ \ref{fig:SF_hierarchy_states}a
and Fig.\@ \ref{fig:SF_hierarchy_states}b and excitations around the modes.

The structure of the matrix element in Eq.\@ (\ref{eq:FF_dsf}) is quite
similar to the matrix element in Eq.\@ (\ref{eq:FF_N}). Thus the dynamical
structure factor exhibits the same hierarchy of excitations (and modes
formed by them) as the spectral function analysed in detail in this
Subsection. The strongest excitations correspond to only a single
electron hole-pair, the first subleading level corresponds to two
electron-hole pairs, and so on.

Also, a similar hierarchy of excitations was observed in numerical
studies of spin chains, \emph{e.g.} Refs. \onlinecite{JS05,JS05_SM,JS07,Biegel02, Biegel03, Takahashi04}.
There, it was found that only a small number of electron-hole pairs
are sufficient to saturate the sum rules for the dynamics response
functions. For example, integration of Eq.\@ (\ref{eq:A_continuum})
over the energy and momentum, 
\begin{equation}
\int d\varepsilon dkA\left(k,\varepsilon\right)=\mathcal{L}-N,\label{eq:A_sumrule}
\end{equation}
gives the number of empty sites. If a sum over only a small number
of electron-hole pairs in the intermediate state $f$ in Eq.\@ (\ref{eq:A_continuum})
is sufficient to fulfil this rule in Eq.\@ (\ref{eq:A_sumrule}) then
a few electron-hole pairs already account for major part of all of
the spectral density and the states with more electron-hole pair have
vanishing spectral weights, as in the hierarchy of modes established in
this work. Our analytic work demonstrates how this can arise in a Bethe Ansatz solution, though 
the numerical studies of spin chains were done at large fillings ($\mathcal{L}\sim N$), for which our result in Eq.\@ (\ref{eq:FF_N}) is not directly applicable.

\subsection{Spectral edge modes}

In this Subsection we consider another important role played by the continuum of eigenstates, namely 
how they form the spectral function close to the spectral edges. These edges separate regions where
there are excitations from regions where there are none,
see borders between white and grey regions in Fig.\@ \ref{fig:SF_spectral_function}. The recently proposed model of a mobile impurity\cite{Khodas06, Khodas072, Imambekov09, Imambekov092} gives
a field-theoretical description of the dynamic response functions
around the spectral edges predicting a general (divergent) power-law
behaviour $A(k,\varepsilon)\sim\left|\varepsilon-\varepsilon_{\textrm{edge}}\left(k\right)\right|^{-\alpha}$, see Refs. \onlinecite{Pereira06, Khodas06, Pereira072, Imambekov09, Imambekov092, Khodas07, Khodas072, Imambekov08, Kamenev09,  Schmidt10, Schmidt102,  Essler10}. For spinless fermions the exponent of
the spectral function is given by\cite{TS13} 
\begin{equation}
\alpha=1-\frac{K}{2}\left(1-\frac{1}{K^{2}}\right)\label{eq:SF_alpha}
\end{equation}
for both the particle ($p0b$) and hole edges ($h0a$), where Eq.
(\ref{eq:SF_Luttinger_params}) gives the Luttinger parameter $K$
in terms of the microscopic parameter of the model in Eq.\@ (\ref{eq:H}).
Here we will compare the field-theoretical result in Eq.\@ (\ref{eq:SF_alpha})
with the microscopic calculation in Eqs. (\ref{eq:kj}, \ref{eq:FF_N}). We find agreement in many cases,
but interestingly we also find some cases where the mobile impurity results are not consistent
with the analytic solution, suggesting this field-theoretical approach is not the complete story.

The hole edge is an $a$-mode, $h0a$, whereas the continuum around
it is dominated by $b$-excitations. The spectral function formed
by these $b$-excitations has already been calculated in Eq.\@ (\ref{eq:Ashape0})
giving the power-law behaviour with the exponent $\alpha=1$. Note
that for spinless fermions the Luttinger parameters have only small
deviations from $K=1$ for arbitrary magnitude of the short-range
interactions. This makes the result in Eq.\@ (\ref{eq:SF_alpha}) $\alpha=1$
for all values of $U$; small deviations (which are $U$-dependent)
make $\alpha<1$ and require subleading terms
in $1/\mathcal{L}$ in the Bethe ansatz calculation in Eqs. (\ref{eq:kj},
\ref{eq:FF_N}). Thus the result of the microscopic calculation coincides
with the prediction of the mobile-impurity model in Eq.\@ (\ref{eq:SF_alpha})
for the hole edge. 
\begin{figure}
\centering{\includegraphics[width=0.95\columnwidth]{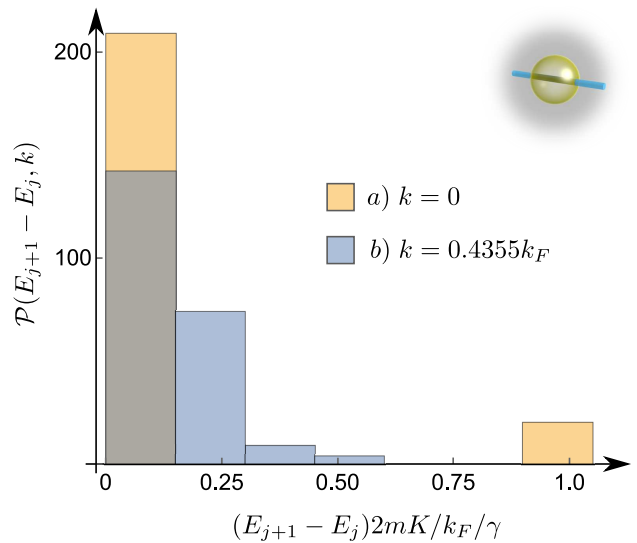}}\protect\caption{\label{fig:SF_edge_level_spacings}Distributions of level spacings in the vicinity ($\textrm{max}\left(E_{f}-\varepsilon_{p0b}\left(k\right)\right)/\varepsilon_{\rm F}=1/100$)
of the particle mode $p0b$ accumulated along energy axis for the
values of momenta a) $k=0$ and b) $k=0.4355k_{\rm F}$; $N=2\cdot10^{3}$
and $L=2\cdot10^{5}$.}
\end{figure}

The particle edge is a $b$-mode, $p0b$, and the excitations around
it belong to the same $b$-level of the hierarchy as the mode itself.
Parameterising the $b$-excitations in this region of the continuum as in Fig.\@ \ref{fig:SF_hierarchy_states}b
and using the the averaging procedure
in Eq.\@ (\ref{eq:average}) we obtain, repeating the same steps as
before, Eqs. (\ref{eq:Ashape0}, \ref{eq:Ashape1}), 
\begin{equation}
\overline{A}\left(\varepsilon\right)\sim\left(\varepsilon-\varepsilon_{p0b}\right)^{3}
\end{equation}
for $k\approx 0$ to 
\begin{equation}
\overline{A}\left(\varepsilon\right)\sim\textrm{const}-\left(\varepsilon-\varepsilon_{p0b}\right)
\end{equation}
for $k\approx k_{\rm F}$, where $\varepsilon_{p0b}\left(k\right)=k_{\rm F}^{2}/\left(mK\right)-k^{2}/\left(2mK\right)$.
This is a new power-law behaviour characterised by an exponent $\alpha$
changing essentially with $k$ from $\alpha=-3$ for $k=0$ to $\alpha=-1$
for $k\approx\pm k_{\rm F}$ and is {\em different} from the predictions of the
mobile-impurity model in Eq.\@ (\ref{eq:SF_alpha}). Here we observe
that the phenomenological model in Refs.\@ \onlinecite{Khodas06,Khodas072, Imambekov09}
is correct only for the $a$-mode spectral edge but higher-order edges
would require a different field-theoretical description.

On a more detailed level, difference between the particle and the
hole edges manifests itself in different statistics of level spacings
around the edges. Evaluation of the density of states, $\nu\left(k,\varepsilon\right)=\sum_{f}\delta\left(\varepsilon-E_{f}\right)\delta\left(k-P_{f}\right)$,
is performed using $E_{f}$ from Eq.\@ (\ref{eq:E_kj}) for a fixed
momentum $k$. For $b$-excitations in Fig.\@ \ref{fig:SF_hierarchy_states}b
we obtain the same results, 
\begin{equation}
\nu\left(k,\varepsilon\right)\sim\left|\varepsilon-\varepsilon_{p0b\left(h0a\right)}\left(k\right)\right|,
\end{equation}
 in the vicinity of both the particle $p0b$ and the hole $h0a$
edges. However the statistics of the level spacings 
\begin{equation}
\mathcal{P}\left(s,k\right)=\sum_{f}\delta\left(s-\left(E_{f+1}-E_{f}\right)\right)\delta\left(k-P_{f}\right),
\end{equation}
where $E_{f}$ are assumed sorted by their values, is different in
the two regions. For the hole edge the energy levels are spaced regularly and are
governed by the slope of dispersion at the Fermi energy $\approx v$. This
gives a bimodal $\mathcal{P}\left(s,k\right)$ with a sharp peak
at $s=0$ (due to many-body degeneracy of almost linear spectrum at
$E_{\rm F}$) and at $s\approx v\gamma$. For the particle
edge the statistics of the level spacings varies from having a regular
level spacing (for $k$ commensurate with $k_{\rm F}$ in Fig.\@ \ref{fig:SF_edge_level_spacings}a) to an irregular distribution (for incommensurate $k$ in Fig.\@ \ref{fig:SF_edge_level_spacings}b). 

The change in the characteristics of the underlying statistics is another
microscopic difference between the particle ($b$-mode) and the hole
($a$-mode) edges that signals a difference in underlying physics for the particles
and for the holes spectral edges beyond the low energy region.

\section{Local density of states}

\begin{figure}
\centering{\includegraphics[width=0.95\columnwidth]{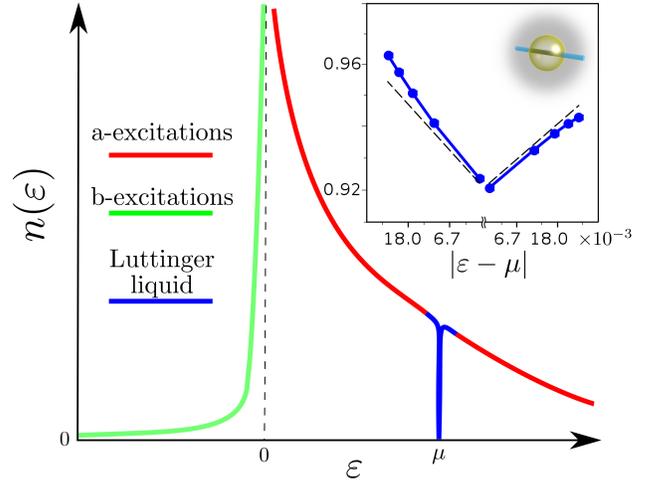}}\protect\caption{\label{fig:SF_LDOS}The local density of states for interacting spinless fermions:
the red and the green lines show the contribution of $a$- and $b$-excitations
and the blue line indicates the Luttinger-liquid regime. Inset is
a log-log plot around the Fermi energy: the points are numerical
data for $N=71$, $L=700$, $mV=6$ giving $K=0.843$, and the dashed
line is $n\left(\varepsilon\right)=\textrm{const}\left|\varepsilon-\mu\right|^{\left(K+K^{-1}\right)/2-1}$.}
\end{figure}
Now we turn to another macroscopic observable, the local density of states
(LDOS), which describes the probability of tunnelling a particle in or
out of the wire at a given position in space and at a given energy.
Since the model in Eq.\@ (\ref{eq:H}) is translationally invariant the
LDOS depends only on a single variable\textendash energy, making
it a more convenient quantity to study qualitatively how the physical
properties change from low to high energies. In this section we will show
how the power-law result of the Tomonaga-Luttinger model\cite{GiamarchiBook}
at low energy crosses over into the hierarchy of modes dominated behaviour
at high energy.

The probability of local tunnelling at energy $\varepsilon$ and at
position $x$ is described by\cite{AGD} $n\left(x,\varepsilon\right)=-\textrm{Im}\big[\int dte^{-i\varepsilon t}G(x,x,t)\big]\textrm{sgn}(\varepsilon-\mu)/\pi$
where $\mu$ is the chemical potential and $G\left(x,x',t\right)=-i\left\langle T\left(e^{-iHt}\psi\left(x\right)e^{iHt}\psi^{\dagger}\left(x'\right)\right)\right\rangle $
is the two point correlation function at zero temperature. In terms
of eigenmodes it reads
\begin{multline}
n\left(\varepsilon\right)=\mathcal{L}\sum_{f}\Big[\left|\left\langle f|\psi^{\dagger}\left(0\right)|0\right\rangle \right|^{2}\delta\left(\varepsilon-E_{f}+E_{0}\right)\\
+\left|\left\langle 0|\psi\left(0\right)|f\right\rangle \right|^{2}\delta\left(\varepsilon+E_{f}-E_{0}\right)\Big].\label{eq:LDOS_def}
\end{multline}
where the coordinate dependence drops out explicitly, the eigenenergy
$E_{f}$ have already been calculated in Eq.(\ref{eq:E_kj}) and the
matrix elements $\left|\left\langle 0|\psi\left(0\right)|f\right\rangle \right|^{2}$
is in Eq.\@ (\ref{eq:FF_N}). Note that the definition in Eq.\@ (\ref{eq:LDOS_def})
is connected to the definition of the spectral function in Eq.\@ (\ref{eq:A_continuum})
via 
\begin{equation}
n\left(\varepsilon\right)=\int dkA\left(k,\varepsilon\right).\label{eq:rho_int_A}
\end{equation}

The leading contribution for $\varepsilon>0$ comes from $a$-excitations.
Substituting the matrix element for the $a$-excitations from Eq.
(\ref{eq:Aa}) we sum over the single-particle-like excitations (with
$\varepsilon=k^{2}/\left(2mK\right)$) that form the mode and obtain
\begin{equation}
n\left(\varepsilon\right)=\sqrt{\frac{2mK}{\varepsilon}}\theta\left(\varepsilon\right).\label{eq:LDOS_a}
\end{equation}
This result gives the same $1/\sqrt{\varepsilon}$ functional dependence \textendash see red line in Fig.\@ \ref{fig:SF_LDOS}\textendash and the same $1/\sqrt{\varepsilon}$ van Hove singularity
at the bottom of the band $\varepsilon=0$ as the free-particle model.

For $\varepsilon<0$ the leading contribution to $n\left(\varepsilon\right)$
comes from $b$-excitations. Instead of performing a summation in
Eq.\@ (\ref{eq:LDOS_def}) over every $b$-excitation in this region
we use an intermediate result in Eq.\@ (\ref{eq:Ashape0}) where the
matrix elements of $b$-excitations in Eq.\@ (\ref{eq:FF_2}) are already
smoothed over many eigenstates and the relation in Eq.\@ (\ref{eq:rho_int_A}).
Evaluating the integral over $k$ in Eq.\@ (\ref{eq:rho_int_A}), after
the substitution of Eq.\@ (\ref{eq:FF_2}) into it, for $\varepsilon<0$ we obtain
\begin{multline}
n\left(\varepsilon\right)=\frac{2Z^{2}k_{\rm F}^{2}}{\gamma\mu K}\theta\left(-\varepsilon\right)\\
\times\left[2\left(1-\frac{3\left|\varepsilon\right|}{\mu}\right)\sqrt{\frac{\mu}{\left|\varepsilon\right|}}\cot^{-1}\left(\sqrt{\frac{\left|\varepsilon\right|}{\mu}}\right)+6\right].\label{eq:rho_b}
\end{multline}
There is a finite probability to find a particle below the bottom
of the conduction band\textendash green line in Fig.\@ \ref{fig:SF_LDOS}\textendash which is allowed only due to interactions between many
particles. The factor $Z$ is proportional to the interaction strength
$V$\textendash see Eq.\@ (\ref{eq:FF_N})\textendash making $n\left(\varepsilon\right)=0$
for $\varepsilon<0$ in the free particle limit of $V=0$. At the bottom
of the band below the $\varepsilon=0$ point in Fig.\@ \ref{fig:SF_LDOS},
the result in Eq.\@ (\ref{eq:rho_b}) contains another Van Hove singularity
\begin{equation}
\rho\left(\varepsilon\right)=\frac{2\pi Z^{2}k_{\rm F}^{2}}{\gamma K\sqrt{\mu\left|\varepsilon\right|}},
\end{equation}
which also disappears when $V=0$. The appearance of the identical exponent as in Eq.\@ (\ref{eq:LDOS_a})
exponent $1/\sqrt{\left|\varepsilon\right|}$ seems to coincidental.

Around the Fermi energy (the point $\varepsilon=\mu$
in Fig.\@ \ref{fig:SF_LDOS}) the Tomonaga-Luttinger
model predicts a power-law suppression of LDOS, 
\begin{equation}
n\left(\varepsilon\right)\sim\left|\varepsilon-\mu\right|^{\left(K+K^{-1}\right)/2-1},\label{eq:LDOS_LL}
\end{equation}
e.g. see the book in Ref.\@ \onlinecite{GiamarchiBook}. However the result
for the $a$-mode in Eq.\@ (\ref{eq:LDOS_a}) is finite at this point,
$n\left(\mu\right)=\sqrt{2mK/\mu}$. In order to resolve the apparent
discrepancy we evaluate $n\left(\varepsilon\right)$ numerically around
the Fermi energy using the determinant representation of the form factors
Eqs. (\ref{eq:psi_1_ABA}, \ref{eq:Mab}) instead of Eq.\@ (\ref{eq:FF_N}),
which accounts for all orders in $1/\mathcal{L}$, and indeed find
a suppression of LDOS around $\varepsilon=\mu$, see blue line in
the inset in Fig.\@ \ref{fig:SF_LDOS}. This  signals that the leading-order
expansion in the $\mathcal{L}\left|\left\langle f|\psi^{\dagger}\left(0\right)|0\right\rangle \right|^{2}=1$
result is insufficient at low energies. Very close to the Fermi point
(the linear region of the single-particle dispersion) all $1/\mathcal{L}$
orders of the Bethe ansatz calculation are needed to reproduce the
result of the Tomonaga-Luttinger model, see dashed lines in the inset in Fig.\@
\ref{fig:SF_LDOS}. However, away from the linear region the particle-hole
symmetry of the Tomonaga-Luttinger model is broken by the finite curvature
of the dispersion and only the leading $1/\mathcal{L}$ order in Eq.
(\ref{eq:FF_N}) is sufficient to account for the main contribution
there.

The general picture emerging in Fig.\@ \ref{fig:SF_LDOS} is a power-law
crossover between different energy scales. At low energies (blue region in Fig.\@ \ref{fig:SF_LDOS}) Eq.\@ (\ref{eq:FF_N})
breaks down and the collective modes of the Tomonaga-Luttinger model
are a better description of the system. At high energies (the red and the green regions in Fig.\@ \ref{fig:SF_LDOS})
the hierarchy of modes, which directly follows from Eqs. (\ref{eq:FF_N},
\ref{eq:kj}), becomes the dominant physical picture. For spinless
fermions the extent of the crossover region is large due to only small
deviations from $K=1$ for arbitrary short-range interactions. For
very small exponents $\left[\left(K+K^{-1}\right)/2-1\right]\ll1$
the power-law in Eq.\@ (\ref{eq:LDOS_LL}) deviates significantly from $1$ only in
an extremely narrow region around $\varepsilon=\mu$ having a large
window of energies where it overlaps with the $a$-mode result in
Eq.\@ (\ref{eq:LDOS_a}).

\section{Experiments on spinful fermions}

So far, in this paper, we have established the theoretical framework
for expecting a hierarchy of modes in a interacting system at high
energy. Now we turn to a measurement of tunnelling of electrons in
a one-dimensional (1D) nanostructure, which gives experimental evidence
for the existence of the hierarchy. Electrons have spin 1/2, which does
not correspond directly to the model of spinless fermions in Eq.\@ (\ref{eq:H}), and there is currently no known method for calculating the necessary form factors for spinful fermions. However, the general picture that emerges from the experiment is qualitatively the same as our result in the theory part of this paper.
\begin{figure}
\includegraphics[width=0.98\columnwidth]{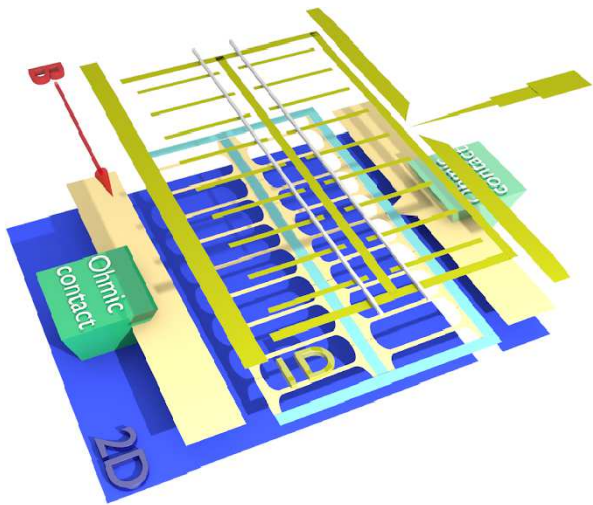}\protect\caption{\label{fig:device} Schematic of the device made out of a double-well heterostructure. The dark blue and cream layers are the lower and upper quantum wells, respectively. The lower layer hosts the two-dimensional electron gas (2DEG). The wires are defined in the upper layer by gating. The gold top layer represents metallic gates deposited on the surface of the semiconductor heterostructure. The array of parallel `finger' gates defines the 1D wires in the upper well. The white lines represent air bridges joining the finger gates together. Current is injected from the ohmic contact on the right solely into the upper well through the constriction at top right in the diagram. The constriction is formed and pinched off by a split pair of gates, and charge is induced again in the upper well in the constriction by a gate in the centre of the channel. The current then flows into the 1D wires via the narrow, nominally 2D, regions shown in light blue. Tunneling to the 2DEG below is possible, and this gives a small `parasitic' current in parallel with the tunnel current from the 1D wires. To tune this tunneling off resonance in the regions of interest by changing the density, a `p' gate is placed above the `p' regions and a voltage $V_{\rm P}$ is applied. Current is prevented from flowing from the upper well into the left-hand ohmic contact by a barrier gate shown on the left, which only depletes the upper well. The red arrow shows the direction of the externally applied magnetic field $B$, which is in the plane of the wells and perpendicular to the wires.}
\end{figure}

The design of our device\cite{Jompol09} is based on a high-mobility
GaAs--AlGaAs double-quantum-well structure (blue and yellow
layers in Fig.\@ \ref{fig:device}), with electron densities around $3$ and $2\times10^{15}$\,m$^{-2}$ in the upper and lower layers, respectively, before application of gate voltages. Electrons in the upper layer are confined to a 1D geometry (`wires') in the $x$-direction by applying a negative voltage to split `finger' gates on the surface (gold layer in Fig.\@ \ref{fig:device}).

Electrons underneath the gates are completely depleted, but electrons below the gap between gates are squeezed into a narrow 1D wire. The extremely regular wires are arranged in an array containing $\sim 600$ of them to boost the signal. The small lithographic width of the wires, $\sim 0.18\,\mu$m, provides a large energy spacing between the first and second 1D subbands defined by spatial modes perpendicular to the wires ($\sim 3-5$\,meV, probably somewhat smaller than for overgrown wires\cite{Yacoby02}). This allows a wide energy window for electronic excitations in the single 1D subband that covers a range of a few chemical potentials of the 1D system. The lower 2DEG (blue in Fig.\@ \ref{fig:device}), is separated from the wires by a $d=14$\,nm tunnel barrier. The wafer is doped with Si symmetrically in the AlGaAs barriers above and below the pair of wells. The doping is separated from the wells by spacer layers. The spacing between the centres of the two quantum wells is nominally $d=32$\,nm but we find a value of $d=35$\,nm fits the data better, and this can be explained by the fact that the centres of the wavefunctions will be slightly further apart owing to the opposite electric fields in each well.

The 2DEG in the lower (dark blue) layer is used as a controllable injector or collector of electrons for the 1D system.\cite{Kardynal96} The current $I$ tunneling in the $z$-direction between the layers is proportional to the convolution of the 1D and 2D spectral functions (a pair of peaks at $k_x=\pm k_{\rm F}$ broadened in $k_y$ by the 1D confinement, and a circle, respectively). An in-plane magnetic field $B$ applied in the $y$-direction, perpendicular to the wires (show with a red arrow in Fig.\@ \ref{fig:device}), produces a Lorentz force that changes the longitudinal momentum $k_x$ acquired while tunneling between layers by $\Delta k=eBd/\hbar$, where $e$ is the electronic charge. Thus $B$ shifts the spectral functions in $k_x$ relative to each other, and so probes the momentum. One spectral function can also be shifted relative to the other in energy by applying a voltage $V$ between the layers, in order to probe the 1D and 2D dispersion relations at different energies. The conductance $G={\rm d}I/{\rm d}V$ has a peak when sharp features in the spectral functions have significant overlap. At $V=0$ this occurs when $\Delta k$ is equal to the sum or difference of the Fermi wavenumbers $k_{\rm F}$ and $k_2$ of the 1D and 2D systems respectively, so there are two peaks for $B>0$, at $B_\pm=\frac{\hbar}{ed}|k_2\pm k_{\rm F}|$. By sweeping $B$ and $V$ one can map out the dispersion relation of states in each layer. The range of magnetic fields that we apply to the system is still within the regime of Pauli paramagnetism for the electron densities in our samples.

\section{Low energy}

\begin{figure}
\includegraphics[width=0.98\columnwidth]{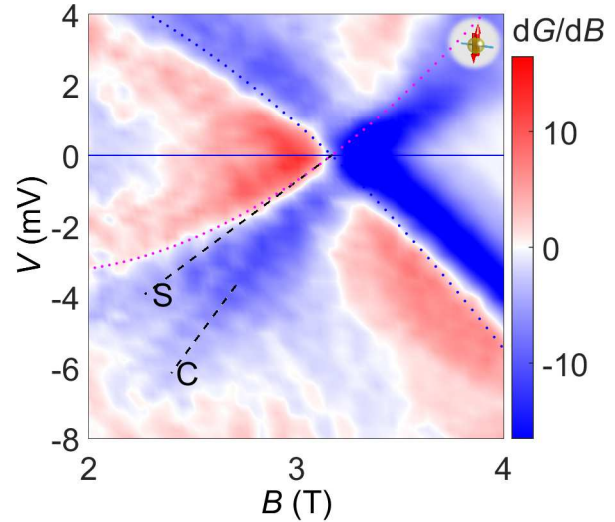}\protect\caption{\label{fig:sc_separation} Intensity plot of ${\rm d}G/{\rm d}B$ at low energies around the Fermi point $k_{\rm F}$. Spin (S) and charge (C) dispersions are indicated by dashed lines. The dotted lines indicate the parabolae expected in the non-interacting model. The finger-gate voltage $V_{\rm F}=-0.70$\,V and the temperature $T\sim300$\,mK.}
\end{figure}


\begin{figure*}[t] 
\includegraphics[width=2\columnwidth]{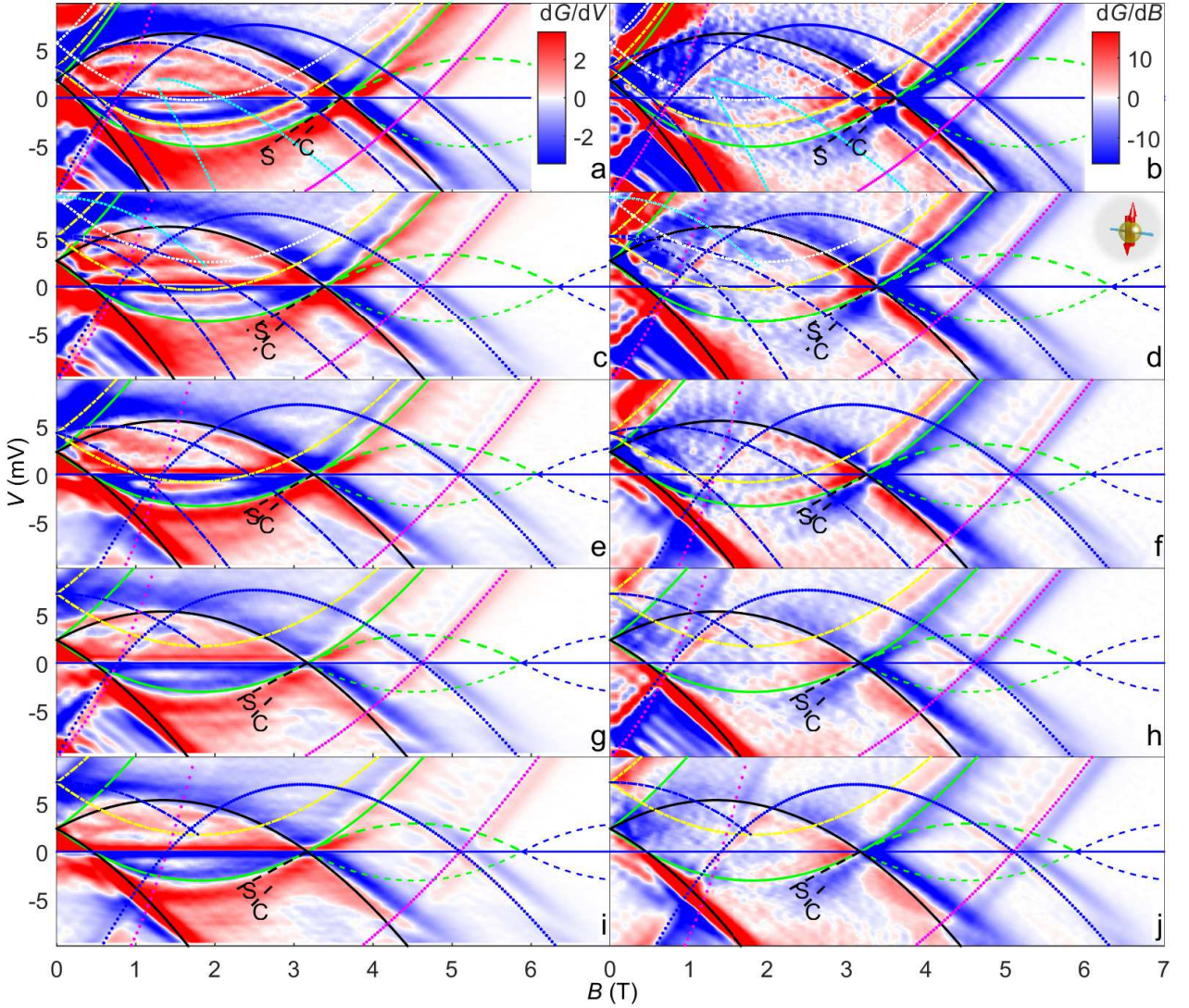}\protect\caption{\label{fig:experiment_main} Intensity plots of ${\rm d}G/{\rm d}V$ (left column, in $\mu$S/mV) and ${\rm d}G/{\rm d}B$ (right column, in $\mu$S/T) from below $-k_{\rm F}$ to above $3k_{\rm F}$ and from $\sim-2\mu$ to $\sim 2\mu$, for various finger-gate voltages: $-0.60$\,V (a, b), $-0.65$\,V (c, d), $-0.68$\,V (e, f) and $-0.70$\,V (g--j). The solid black lines map out the dispersion of the lower (2D) layer. The green solid line marks $a$-modes, thick and thin dashed green lines, $p1b$ and $h1b$ modes, respectively, and dashed blue, higher-$k$ modes (as in Fig.\@ \ref{fig:SF_spectral_function}). Dot-dashed yellow (blue) and dotted white (cyan) lines show second and third 1D subbands (2D dispersion measured by those subbands), respectively (though the third is empty, electrons can tunnel into it from the 2D layer and hence there are sometimes signs of its effects for $V>0$, especially near $B=0$). Dotted magenta and blue lines are `parasitic' 2D dispersions of the two layers. The voltage on the gate over this region $V_{\rm P}=0$\,V except for e, f ($V_{\rm P}=0.2$\,V) and i, j ($V_{\rm P}=0.3$\,V), which shifts the parabolae to the right without changing the signal from the 1D wires. The lines have all been adjusted to take account of the capacitive coupling between the layers.
Spin (S) and charge (C) modes are indicated with black dashed lines. $T\sim300$\,mK. See Table \ref{tab:vcvs_ratio} for the densities and the ratio $v_{{\rm c}}/v_{{\rm s}}$ for each gate voltage.}
\end{figure*}

\begin{table}
\begin{ruledtabular}
\begin{tabular}{c|c|c|c|c}
$V_\textrm{F}$ (V) & $n_{\rm 2D} (10^{15}\,{\rm m}^{-2})$ & $n_{\rm 1D} (10^7\,{\rm m}^{-1})$ & $k_{\rm F} (10^7\,{\rm m}^{-1})$ & $v_c/v_s$\tabularnewline
\hline
$-0.60$ & 1.67 & 5.68 & 8.9 & $1.5$ \tabularnewline
\hline
$-0.65$ & 1.65 & 4.99 & 7.8 & $1.6$ \tabularnewline
\hline
$-0.68$ & 1.52 & 4.79 & 7.5 & $1.5$ \tabularnewline
\hline
$-0.70$ & 1.48 & 4.60 & 7.2 & $1.8$ \tabularnewline
\end{tabular}
\end{ruledtabular}
\caption{\label{tab:vcvs_ratio} Densities of the 2D layer ($n_{\rm 2D}$) and of the 1D wires ($n_{\rm 1D}$), the 1D Fermi wavevector $k_{\rm F}$ (all to about $\pm 1\%$), and the ratio of the charge and the spin velocities at low energies (to about $\pm 5\%$), extracted from the gradients of the S and C lines in Fig. \ref{fig:experiment_main}, for different finger-gate voltages $V_\textrm{F}$.}
\end{table}
First, we measure the tunnelling conductance $G={\rm d}I/{\rm d}B$ in a small range of voltages and magnetic fields around $V=0$ and $B=B_+=3.15$\,T that corresponds to a region on the momentum-energy plane around the Fermi point ($\varepsilon=\mu, k=k_{\rm F}$), see Fig.\@ \ref{fig:sc_separation}. Below the Fermi energy we observe splitting of the single-particle line into two with different dispersions---spin (S) and charge (C) separation\cite{Jompol09,Yacoby02}---giving two different slopes $v_{\rm s}$ and $v_{\rm c}$ (black dashed lines in Fig.\@ \ref{fig:sc_separation}). We assume that $v_{\rm s}$ is the same as for non-interacting electrons and so take it to be the gradient of the parabola at $V=0$. We estimate $v_{\rm c}$ from the positions of steepest gradient and hence obtain $v_{\rm s}\approx1.2\times10^{5}$\,ms$^{-1}$ and $v_{\rm c}\approx2.3\times10^{5}$\,ms$^{-1}$ at the finger-gate voltage $V_{\rm F}=-0.70$\,V.

Theoretically, the low-energy physics of the interacting 1D electrons is described well by a spinful generalisation of the Luttinger-liquid model.\cite{GiamarchiBook} Its excitations are collective hydrodynamic-like modes that are split into charge-only and spin-only excitations. For any finite strength of the interactions between fermions the two types of modes have linear dispersions with different slopes $v_{\rm c}$ and $v_{\rm s}$. In the absence of interactions the difference between the two velocities vanishes in accordance with the free-electron model, in which the spin degree of freedom does not affect the spectrum but results only in the double degeneracy of the fermionic states. Thus the ratio of $v_{\rm c}/v_{\rm s}$ serves as a good measure of the interaction strength. Since the Coulomb interaction between electrons is repulsive the charge branch always has a steeper slope $v_{\rm c}\geq v_{\rm s}$ (see Ref.\@ \onlinecite{GiamarchiBook}). Thus the ratio varies from $1$ for free to $\infty$ for infinitely repelling particles. In our experiment we measure the tunneling of electrons and observe two peaks that we attribute to the charge and the spin dispersions. The pair of velocities above gives a large $v_{{\rm c}}/v_{{\rm s}}\approx1.8\pm0.1$ (for $V_\textrm{F}=-0.70$\,V), confirming that our system is in the strongly interacting regime.

\section{High energy}

Now we extend the ranges of the voltage and magnetic field measuring the tunneling conductance $G$ across the double quantum well in Fig.\@ \ref{fig:device} accessing a large portion of the 1D spectral function from below $-k_{\rm F}$ to $3k_{\rm F}$ and from $-2\mu$ to $2\mu$, see Fig.\@ \ref{fig:experiment_main}. There is an unavoidable `parasitic' (`p') tunneling from narrow 2D regions (light blue strips in Fig.\@ \ref{fig:device}) that connect the wires to the injector constriction. This superimposes a set of parabolic 2D-2D dispersions on top of the 1D-2D signal, which are marked by magenta and blue dotted lines in Fig.\@ \ref{fig:experiment_main}. 

Apart from the parasitic and the 2D dispersion signals, we observe only a single 1D parabola away from $B=0$, marked by the solid green line in Fig.\@ \ref{fig:experiment_main}. It extends from the spin-excitation branch at low energy and the position of its minimum multiplied by the electronic charge $e$ gives the 1D chemical potential $\mu\approx 4$\,meV. The $B_-$ and $B_+$ crossings with the line $V=0$, corresponding to momenta $-k_{\rm F}$ and $k_{\rm F}$, give the 1D Fermi momentum $k_{\rm F}\approx8\times10^{7}$\,m$^{-1}$. All other edges of the 1D spectral function are constructed by mirroring and translation of the hole part of the observable 1D dispersion, dashed green and blue lines in Fig.\@ \ref{fig:experiment_main}. 

\begin{figure}[t] 
\includegraphics[width=1\columnwidth]{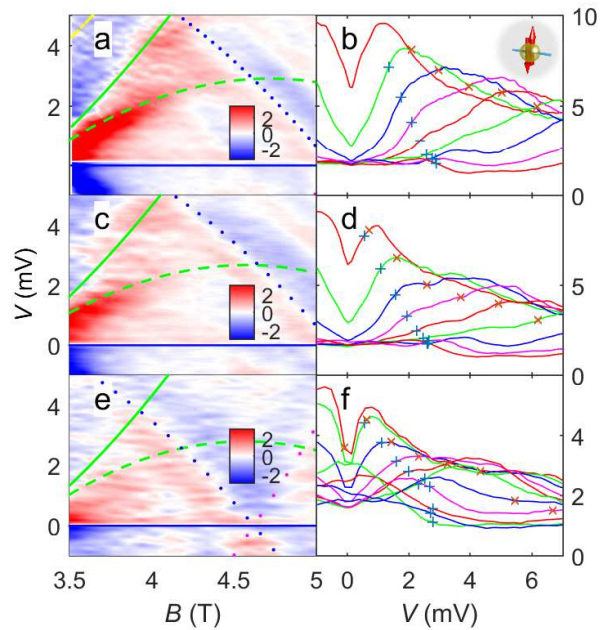}\protect\caption{\label{fig:p1b} Left column: intensity plots of ${\rm d}G/{\rm d}V$ (in $\mu$S/mV), for various finger-gate voltages and samples: (a, b) $V_{\rm F}=-0.68$\,V, $V_{\rm P}=0.2$\,V, from Fig.\@ \ref{fig:experiment_main}e, sample A, which had $10\,\mu$m-long wires ($T\sim300$\,mK); (c, d) $V_{\rm F}=-0.70$\,V, $V_{\rm P}=0.3$\,V, from Fig.\@ \ref{fig:experiment_main}i, sample A; (e, f) a similar single-subband result from sample B ($18\,\mu$m-long wires, $T<100$\,mK). The replica feature just above $k_{\rm F}$ appears as a pale triangle (slowly varying $G$) between the two green curves, after a red region (sharp rise in $G$). The replica feature for sample B is somewhat weaker than that for sample A, in line with the wire-length dependence predicted in this paper. Right column: $G$ \textit{vs} $V$ at various fields $B$ from 3 to 4.8\,T for the data in the matching plots in the left column; `+' and `$\times$' symbols on each curve indicate, respectively, the voltages corresponding to the dashed and solid ($p1b$ and $p1a(l)$) green lines in the left column (and in Fig.\@ \ref{fig:experiment_main}), showing the enhanced conductance between the two.}
\end{figure}
For positive voltages in the region just above the higher $V=0$ crossing point ($B_+$, which corresponds to $k_{\rm F}$) we observe a distinctive feature: the 1D peak broadens, instead of just continuing along the non-interacting parabola, with one boundary following the parabola ($p1a(l)$) and the other bending around, analogous to the replica $p1b$. This is visible in the conductance, but is most easily seen in the differentials, particularly ${\rm d}G/{\rm d}V$ (left column of Fig.\@ \ref{fig:experiment_main}). The broadening is observed at temperatures from 100\,mK up to at least 1.5\,K, and in samples with different wire designs (with or without air bridges) and lengths: in Fig.\@ \ref{fig:p1b}, ${\rm d}G/{\rm d}V$ is shown in detail for the broadened `replica' region for the 10\,$\mu$m wires already presented (a--d), and for another sample with wires 18\,$\mu$m long (e, f). $G$ is plotted in Fig.\@ \ref{fig:p1b}b, d and f on cuts along the $V$ axis of the corresponding plots in the left column at various fields $B$ from $B_+$ to 4.8\,T---between the `+' and `$\times$' symbols on each curve is a region of enhanced conductance characteristic of the replica $p1b$.

Filling of the second 1D subband changes significantly the screening radius for the Coulomb interaction potential in the first 1D subband. This is manifested by a change of the ratio $v_{\rm c}/v_{\rm s}$ when the occupation of the second subband is changed by varying voltage of the finger gates $V_\textrm{F}$ in Figs. \ref{fig:experiment_main}aceg, see Table \ref{tab:vcvs_ratio}. The ratio $v_{\rm c}/v_{\rm s}$ is a measure of interaction energy. Thus, the finger gates give a degree of experimental control over the interactions within our design of the 1D system. We use the maximum change of the ratio $v_c/v_s$  for different finger gate voltages to estimate the relative change of the interaction strength as $\left(\textrm{max}(v_c/v_s)-\textrm{min}(v_c/v_s)\right)/\textrm{min}(v_c/v_s)$ obtaining a change of about $20\%$.

It also has to be noted that the `replica' is visible even when a second subband is present in the 1D wires, see Fig. \ref{fig:experiment_main}a--f. In a and b it appears to go 25--30\% higher in voltage than expected for a precise copy of the usual 1D parabola (even allowing for capacitive correction) due to a contribution of the second subband, which we do not analyse in detail here.


At even higher magnetic fields the $p1b$ line passes a `p' parabola. Figs.\@ \ref{fig:p1b}a and c (and the corresponding cuts b and d) show the replica feature for two different positions of the `p' parabolae using a gate above most of the `p' region, showing that the replica feature is independent of the `p' tunneling. The amplitude of the feature dies away rapidly, and beyond the `p' parabolae, we have measured up to 8T with high sensitivity, but find no sign of any feature that can be distinguished from the decaying tails of the other features.

In the range of fields where the $p1b$ feature is observed its strength decreases as the $B$ field increases away from the crossing point analogously to the power-law for spinless fermions in Table\@ \ref{tab:SF_spectral_function_values}. On general grounds it is natural to expect that divergence of the spectral weight of a b-mode toward an a-mode is a general feature, but there is no known method for performing a microscopic calculation in the spinful case. A similar feature should mark the $h0b\left(r\right)$ mode (see Fig.\@ \ref{fig:SF_spectral_function} and Table \ref{tab:SF_spectral_function_values}) for negative voltages and for the magnetic field just below the crossing point $k_{\rm F}$, but it would be very difficult to resolve due to the overlaying spin and charge lines.

\begin{figure}\includegraphics[width=1\columnwidth]{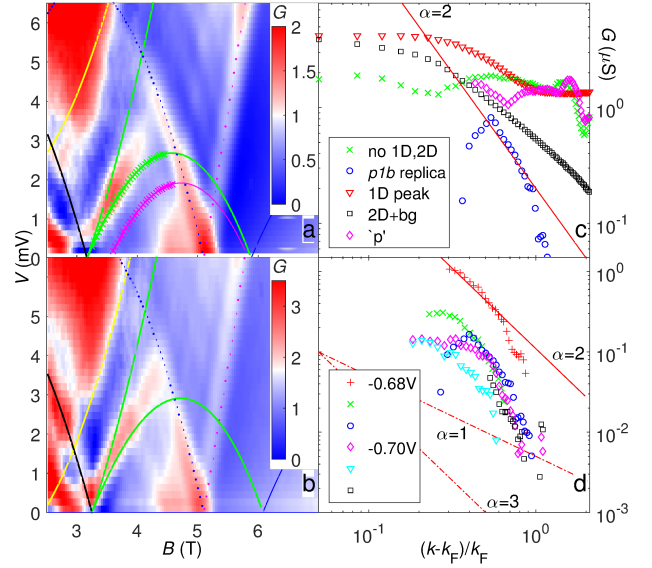}\protect\caption{\label{fig:replicafitting} (a,c) The conductance for $V_{\rm F}=-0.70$ and $-0.68\,V$, respectively, after subtraction of an idealised landscape made up of fits or estimates of the non-interacting 1D-2D and `p' parabolae (see text). The $p1b$ replica is seen clearly as the red region of enhanced conductance. (b) The conductance along the $p1b$ replica parabola, for the data in (a) (green crosses). The conductance on $p1b$ has a large contribution from the `p' region (the line in (a) marked with blue dots, which is blurred to the left by multiple copies at slightly different positions). In order to correct for this contribution, the conductance along a matching parabola shifted along the dotted  `p' line in (a) (shown as a dashed magenta line there), is subtracted from the $p1b$ data. This yields the points marked with blue circles, which appears to be non-zero because of the enhancement at $p1b$. The amplitude decays rapidly. There are many uncertainties in the fitting of the other peaks, but the replica appears clearly and the decay of the conductance is consistent with an inverse-square power law $G\propto (k-k_{\rm F})^{-\alpha}$ (labelled $\alpha=2$), which is the behaviour predicted by the theory for $k>k_{\rm F}+\gamma$ where $\gamma\ll k_{\rm F}$ (see Table \ref{tab:SF_spectral_function_values}). (d) The $p1b$ conductance enhancement as shown with circles in (c). Three different methods of fitting the background and the 1D and 2D peaks are compared for each of two gate voltages as shown. The curves are offset vertically for clarity. The lines marked with values of $\alpha$ are guides to the eye. The data are all consistent with $\alpha=2\pm1$.}
\end{figure}
Making an analogy with the microscopic theory for spinless fermions in the first part of this paper, we estimate the ratio of signals around different spectral edges using the 1D Fermi wavelength, $\lambda_{\rm F}\approx130$\,nm for our samples, as the short-range scale $\mathcal{R}$. The signal from the principal parabola, see Fig.\@ \ref{fig:experiment_main}b, gives the amplitude of the $a$-mode as $G_{a}\approx 5\,\mu$S. Then the amplitude of the signal from the second (third)-level excitations is predicted to be smaller by a factor of more than $\lambda_{\rm F}^{2}/L^{2}\sim 2\times10^{-4}$ ($\lambda_{\rm F}^{4}/L^{4}=3\times10^{-8}$), where the length of a wire is $L=10\,\mu$m. These values $G_{a}\lambda_{\rm F}^{2}/L^{2}\sim 10^{-3}\,\mu$S ($G_{a}\lambda_{\rm F}^{4}/L^{4}\sim 10^{-7}\,\mu$S) are at least two orders of magnitude smaller than the background and noise levels of our experiment $G_{\textrm{noise}}\sim 10^{-2}\,\mu$S, which places an upper limit on the amplitude of any replica away from $k_{\rm F}$. Thus, our observations are consistent with the mode hierarchy picture for fermions.

In an effort to quantify the decay of the replica feature we have fitted the gradual background fall in conductance and the non-interacting 1D and 2D peaks (solid green and blue lines in Figs.\@ \ref{fig:sc_separation}--\ref{fig:p1b}) with a Gaussian and/or Lorentzian functions of $B$, at each value of $V>0$. The fitting parameters are then fitted to smooth functions in order to represent the general behaviour of the peaks as a function of $V$. This idealised landscape is then subtracted from the data, see Fig.\@ \ref{fig:replicafitting}a, and the `replica' is then fairly easily observed in the remaining conductance. A copy of a nearby region along the `p' curve is then subtracted too, as an approximation to the rather diffuse signal arising from the main `p' peak and smaller versions of it at slightly different densities. This also reduces errors in the peak and background fitting used in (a). We then plot the conductance along the expected parabola (dashed line in (a)) as a function of $(k-k_{\rm F})/k_{\rm F}=(B-B_+)/((B_+-B_-)/2)$. This data is shown as circles in (c), where all the other contributions to the conductance along the same parabola are shown. Here, $B_+=3.17$\,T and $k_{\rm F}=7.2\times 10^7$\,m$^{-1}$. It is very hard to be sure that this procedure is reliable due to significant error bars imposed by contributions from the various other peaks, but it is clear that the replica feature dies away rapidly as a
function of $k-k_\textrm{F}$, and it is consistent with the $1/(k-k_\textrm{F})^2$ law predicted for $p1b$ in Table \ref{tab:SF_spectral_function_values} for $k-k_{\rm F}\gg \gamma$. Though the overall prefactor is unknown theoretically in the spinful case, this singular power law may overcome the reduction factor $\mathcal{R}^2/L^2$ close to $k_\textrm{F}$.


\section{Conclusions}

In this work, we have shown that a hierarchy of modes emerges in systems of interacting fermions in one dimension at high energy controlled by the system length, in marked contrast to the well-known fermionic quasiparticles of a Fermi liquid and hydrodynamic modes of a Luttinger liquid at low energy.

We have obtained theoretically the dynamic response functions for a model of spinless fermions with short-range interactions using the exact diagonalisation methods of the Bethe ansatz for the spectrum and the form factors of the system. Analysing the spectral function in detail, we have found that the first-level (strongest) mode in long systems has a parabolic dispersion, like that of a renormalised free particle. The second-level excitations produce a singular power-law line shape for the first-level mode and different kinds of power-law behaviour at the spectral edges. Evaluating the form factor necessary for the dynamical structure factor we have shown that it has the same general form as the form factor of the spectral function, manifesting the same hierarchy of modes.

Using the same many-body matrix elements obtained microscopically, we have also calculated the local density of states. It provides a more convenient way to analyse how the hierarchy at high energy changes into the hydrodynamic modes of the Luttinger liquid at low energies. We have shown, via a full Bethe-ansatz calculation, that the LDOS is suppressed at the Fermi energy in a power-law fashion in full accord with the prediction of the Tomonaga-Luttinger model. Away from the Fermi point, where the Lorentz invariance of the linear dispersion is reduced to Galilean by the parabolicity of the spectrum, the LDOS is dominated by the first (leading) level of the hierarchy. We have demonstrated that the transition from one regime to another is a smooth cross-over.

We measure momentum-resolved tunnelling conductance in one-dimensional wires formed in the GaAs/AlGaAs double-well heterostructure by an array of finger gates. In this set-up we probe the spectral function of unpolarised electrons (spinful fermions) and find a pronounced spin-charge separation at low energy with a ratio of the spin and the charge velocities up to 1.8, which confirms that our system is in the strongly interacting regime. By varying the gate voltage that controls the width of our 1D wires, we demonstrate control of the interaction strength of about $20\%$; the deeper confining potential of the wires populates higher 1D subbands as well which in turn screens stronger Coulomb interactions in the principal 1D band reducing the interaction strength. In $ 10\,\mu$m-long wires we find a clear feature resembling the second-level excitations, which dies away rapidly at high momentum. A qualitative fit shows that the feature decays in a fashion that is consistent with the power-law prediction in this paper for spinless electrons.

Thus we have shown that the hierarchy is apparently a generic phenomenon at least for one- and two-point correlation functions of fermions without spin, and for a transport experiment for fermions with spin.


\begin{acknowledgments}
We acknowledge financial support from the UK EPSRC through grant numbers EP/J01690X/1 and EP/J016888/1 and from the  DFG though SFB/TRR 49. This research was supported in part by the National Science Foundation under Grant No. NSF PHY11-25915. 
\end{acknowledgments}

\appendix
\onecolumngrid

\section{Eigenvalue equation in the algebraic framework}

The eigenvalue of the transfer matrix $\tau(u)$ in Eq.\@ (\ref{eq:transfer_matrix})
can be evaluated using the commutation relations in Eqs. (\ref{eq:AC},\ref{eq:DC}).
Let $\Psi$ be a Bethe state in the algebraic representation of Eq.
(\ref{eq:psiN_algebraic}). The results of acting with $A\left(u\right)$
and $D\left(u\right)$ operators on the state $\Psi$ are obtained by commuting these
operators from left to right though the product of $C\left(u_{j}\right)$
in Eq.\@ (\ref{eq:psiN_algebraic}) and then by using the their vacuum
eigenvalues in Eq.\@ (\ref{eq:vacuum_ad}). 

Let us consider the case of $N=2$ and the operator
$A\left(u\right)$ first. Commuting once by means of Eq.\@ (\ref{eq:AC})
gives
\begin{equation}
A\left(u\right)C\left(u_{2}\right)C\left(u_{1}\right)\left|0\right\rangle =\left(\frac{1}{b\left(u_{2}-u\right)}C\left(u_{2}\right)A\left(u\right)-\frac{c\left(u_{2}-u\right)}{b\left(u_{2}-u\right)}C\left(u\right)A\left(u_{2}\right)\right)C\left(u_{1}\right)\left|0\right\rangle .
\end{equation}
Applying Eq.\@ (\ref{eq:AC})  the second time gives
\begin{eqnarray}
A\left(u\right)C\left(u_{2}\right)C\left(u_{1}\right)\left|0\right\rangle  & = & \left(\frac{1}{b\left(u_{2}-u\right)}\frac{1}{b\left(u_{1}-u\right)}C\left(u_{2}\right)C\left(u_{1}\right)a\left(u\right)-\frac{c\left(u_{1}-u\right)}{b\left(u_{1}-u\right)}\frac{1}{b\left(u_{2}-u\right)}C\left(u_{2}\right)C\left(u\right)a\left(u_{1}\right)\right)\left|0\right\rangle \\
 &  & +\left(-\frac{1}{b\left(u_{1}-u_{2}\right)}\frac{c\left(u_{2}-u\right)}{b\left(u_{2}-u\right)}C\left(u\right)C\left(u_{1}\right)a\left(u_{2}\right)+\frac{c\left(u_{1}-u_{2}\right)}{b\left(u_{1}-u_{2}\right)}\frac{c\left(u_{2}-u\right)}{b\left(u_{2}-u\right)}C\left(u\right)C\left(u_{2}\right)a\left(u_{1}\right)\right)\left|0\right\rangle ,\nonumber 
\end{eqnarray}
where the vacuum eigenvalue of $A\left(u\right)$, $A\left(u\right)\left|0\right\rangle =a\left(u\right)\left|0\right\rangle $,
was substituted explicitly. The second terms in the first and the second
lines of the above equation have the same operator form but different
scalar factors. Summation of the two scalar factor, using the explicit
form of $b\left(u\right)$ and $c\left(u\right)$ from Eq.\@ (\ref{eq:bc_def}),
yields
\begin{equation}
\frac{a\left(u_{1}\right)}{b\left(u_{2}-u\right)}\left[\frac{c\left(u_{1}-u_{2}\right)}{b\left(u_{1}-u_{2}\right)}c\left(u_{2}-u\right)-\frac{c\left(u_{1}-u\right)}{b\left(u_{1}-u\right)}\right]=-\frac{a\left(u_{1}\right)}{b\left(u_{2}-u_{1}\right)}\frac{c\left(u_{1}-u\right)}{b\left(u_{1}-u\right)}.
\end{equation}
Thus the four terms can be rewritten as only three terms,
\begin{equation}
A\left(u\right)C\left(u_{2}\right)C\left(u_{1}\right)\left|0\right\rangle =\left(a\left(u\right)\prod_{j=1}^{2}\frac{C\left(u_{j}\right)}{b\left(u_{j}-u\right)}-\sum_{j=1}^{2}a\left(u_{j}\right)\frac{c\left(u_{j}-u\right)}{b\left(u_{j}-u\right)}C\left(u\right)\prod_{l=1\neq j}^{2}\frac{C\left(u_{l}\right)}{b\left(u_{l}-u_{j}\right)}\right)\left|0\right\rangle. 
\end{equation}
Extension of the same procedure for $N>2$ gives
\begin{equation}
A\left(u\right)\prod_{j=1}^{N}C\left(u_{j}\right)\left|0\right\rangle =\left(a\left(u\right)\prod_{j=1}^{N}\frac{1}{b\left(u_{j}-u\right)}C\left(u_{j}\right)-\sum_{j=1}^{N}a\left(u_{j}\right)\frac{c\left(u_{j}-u\right)}{b\left(u_{j}-u\right)}C\left(u\right)\prod_{l=1\neq j}^{N}\frac{1}{b\left(u_{l}-u_{j}\right)}C\left(u_{l}\right)\right)\left|0\right\rangle \label{eq:A_psi}
\end{equation}
Commuting of the operator $D\left(u\right)$ is done in 
the same way using Eq.\@ (\ref{eq:DC}) and yields
\begin{equation}
D\left(u\right)\prod_{j=1}^{N}C\left(u_{j}\right)\left|0\right\rangle =\left(d\left(u\right)\prod_{j=1}^{N}\frac{-1}{b\left(u-u_{j}\right)}C\left(u_{j}\right)+\sum_{j=1}^{N}d\left(u_{j}\right)\frac{c\left(u-u_{j}\right)}{b\left(u-u_{j}\right)}C\left(u\right)\prod_{l=1\neq j}^{N}\frac{-1}{b\left(u_{j}-u_{l}\right)}C\left(u_{l}\right)\right)\left|0\right\rangle .\label{eq:D_psi}
\end{equation}

Thus a Bethe state in Eq.\@ (\ref{eq:psiN_algebraic}) parametrised
by an arbitrary set of $u_{j}$ is not an eigenstate of the transfer matrix
since acting of the operator $\tau(u)$ on such a state does not only
result in multiplying by a scalar but also generates many different states:
the second terms in Eq.\@ (\ref{eq:A_psi}) and in Eq.\@ (\ref{eq:D_psi}).
However, the coefficients in front of each of these extra states can
be made zero by a specific choice of $u_{j}$,
\begin{equation}
a\left(u_{j}\right)\prod_{l=1\neq j}^{N}\frac{1}{b\left(u_{l}-u_{j}\right)}-d\left(u_{j}\right)\prod_{l=1\neq j}^{N}\frac{-1}{b\left(u_{j}-u_{l}\right)}=0.\label{eq:appendix_BE}
\end{equation}
When a set of $u_{j}$ satisfies the system of equations above, the corresponding
Bethe state is an eigenstate of the transfer matrix, $\tau(u)\Psi=\mathcal{T}(u)\Psi$,
with the eigenvalue $\mathcal{T}$ given by the first terms in Eq.
(\ref{eq:A_psi}) and in Eq.\@ (\ref{eq:D_psi}),
\begin{equation}
\mathcal{T}(u)=a\left(u\right)\prod_{j=1}^{N}\frac{1}{b\left(u_{j}-u\right)}-d\left(u\right)\prod_{j=1}^{N}\frac{-1}{b\left(u-u_{j}\right)}.\label{eq:appendix_Tau}
\end{equation}
Eq.\@ (\ref{eq:appendix_BE}) is the set of Bethe equations in the algebraic
representation, Eq.\@ (\ref{eq:BAequation_ABA}) of the main part of
the text, and Eq.\@ (\ref{eq:appendix_Tau}) gives the eigenvalue of
the transfer matrix, Eq.\@ (\ref{eq:tau_eigenvalue}) of the main part
of the text.

\section{Calculation of averages of the local density operator $\rho\left(0\right)$ }

The calculation of the average of the local density operator $\rho\left(0\right)$
is done in the same way as for the field operator $\psi^{\dagger}\left(0\right)$
in Subsection IVc. We start from the lattice model in Eq.\@ (\ref{eq:H_lattice})
and the corresponding construction of the algebraic Bethe ansatz in
Subsection IVa. 

The local density operator can be represented in terms of $A$ and
$D$ operators as\cite{Kitaine99, Kitaine00} 
\begin{equation}
\rho_{1}=-D\left(\frac{i\pi}{2}-\eta\right)\tau\left(\frac{i\pi}{2}-\eta\right)^{\mathcal{L}-1}.
\end{equation}
The action of the second factor in the above expression on an eigenstate
$\left|\mathbf{u}\right\rangle $ just gives a phase factor \textendash{}
see an explanation after Eq.\@ (\ref{eq:psij_aba}) \textendash{} that
we will ignore since we are interested in the modulus squared of this
form factor. Then commuting  the operator $D$ of the first factor
in the equation above through all $C$ operators of the eigenstate
$\left|\mathbf{u}\right\rangle $ \textendash{} in the form in Eq.
(\ref{eq:psiN_algebraic}) \textendash{} gives the result in Eq.\@ (\ref{eq:D_psi}).

The scalar product of Eq.\@ (\ref{eq:D_psi}), where the auxiliary parameter $u$
is set to $i\pi/2-\eta$, with another eigenstate $\left\langle \mathbf{v}\right|$
gives
\begin{multline}
\left\langle \mathbf{v}|\rho_{1}|\mathbf{u}\right\rangle =\left(-1\right)^{N}\prod_{j=1}^{N}\frac{\cosh\left(u_{j}-\eta\right)}{\cosh\left(u_{j}+\eta\right)}\left\langle \mathbf{u}|\mathbf{v}\right\rangle \\
+i\left(-1\right)^{N}\sum_{b=1}^{N}\frac{\sinh2\eta}{\cosh\left(u_{b}+\eta\right)}\prod_{l=1\neq b}^{N}\frac{\sinh\left(u_{b}-u_{l}+2\eta\right)}{\sinh\left(u_{b}-u_{l}\right)}\left\langle u_{b-1},\frac{i\pi}{2}-\eta,u_{b+1}|\mathbf{v}\right\rangle,  \label{eq:psid_psi_1}
\end{multline}
where $\left\langle u_{b-1},\frac{i\pi}{2}-\eta,u_{b+1}\right|$ is
a Bethe state which is constructed from the eigenstate $\mathbf{u}$
by replacing $b^{th}$ quasimomenta with $i\pi/2-\eta$. Note that
the properties $\left\langle \mathbf{v}|\mathbf{u}\right\rangle =\left\langle \mathbf{u}|\mathbf{v}\right\rangle $,
where $u_{j}$ satisfy the Bethe equations and $v_{j}$ is an arbitrary
set of quasimomenta,\cite{Kitaine99, Kitaine00} was used. 

The scalar product in the first line in Eq.\@ (\ref{eq:psid_psi_1})
is given by Eqs. (\ref{eq:scalar_product}, \ref{eq:scalar_product_matrix_elements})
where $\mathbf{u}$ and $\mathbf{v}$ are swapped. Substitution of
$\mathbf{u}=u_{b-1},\frac{i\pi}{2}-\eta,u_{b+1}$ in the same expressions
for the scalar products in the second line of Eq.\@ (\ref{eq:psid_psi_1})
yields
\begin{equation}
\left\langle u_{b-1},\frac{i\pi}{2}-\eta,u_{b+1}|\mathbf{v}\right\rangle =-i\frac{\sinh^{N}\left(2\eta\right)\prod_{j=1}^{N}\cosh\left(v_{j}-\eta\right)\det\hat{T}^{\left(b\right)}}{\prod_{j<i}\sinh\left(v_{j}-v_{i}\right)\prod_{j<i\neq b}\sinh\left(u_{j}-u_{i}\right)\left(-1\right)^{b-1}\prod_{j=1\neq b}^{N}\cosh\left(u_{j}+\eta\right)}
\end{equation}
where all matrix elements of $\hat{T}^{\left(b\right)}$ are 
\begin{multline}
T_{ab'}^{\left(b\right)}=\prod_{l=1\neq b'}^{N}\frac{\sinh\left(u_{b'}-u_{l}+2\eta\right)}{\sinh\left(u_{b'}-u_{l}-2\eta\right)}\frac{1}{\sinh\left(u_{b'}-v_{a}\right)}\prod_{j=1\neq a}^{N}\sinh\left(u_{b'}-v_{j}-2\eta\right)\\
-\frac{1}{\sinh\left(u_{b'}-v_{a}\right)}\prod_{j=1\neq a}^{N}\sinh\left(u_{b'}-v_{j}+2\eta\right)
\end{multline}
for $b'\neq b$
\begin{equation}
\hat{T}_{ab'}^{\left(b\right)}=\frac{1}{\cosh\left(v_{a}+\eta\right)\cosh\left(v_{a}-\eta\right)}
\end{equation}
for $b'=b$.

After pulling a common factor out of the brackets in Eq.\@ (\ref{eq:psid_psi_1})
and absorbing the $b$-dependent prefactors in front of the determinants
in the second line of Eq.\@ (\ref{eq:psid_psi_1}) into the $b^{th}$
columns of the matrices under the determinants, the form factor in
Eq.\@ (\ref{eq:psid_psi_1}) reads as 
\begin{equation}
\left\langle \mathbf{v}|\rho_{1}|\mathbf{u}\right\rangle =\prod_{j=1}^{N}\frac{\cosh\left(u_{j}-\eta\right)}{\cosh\left(u_{j}+\eta\right)}\frac{\sinh^{N}\left(2\eta\right)}{\prod_{j<i}\sinh\left(v_{j}-v_{i}\right)\prod_{j<i}\sinh\left(u_{j}-u_{i}\right)}\left[\det\hat{T}+\sum_{b=1}^{N}\det\hat{\tilde{T}}^{\left(b\right)}\right]\label{psid_psi_1_sum_of_dets}.
\end{equation}
Here the matrix elements of $\hat{\tilde{T}}^{\left(b\right)}$,
which are obtain by multiplying by the corresponding scalars, are
\begin{multline}
\tilde{T}_{ab'}^{(b)}=\prod_{l=1\neq b'}^{N}\frac{\sinh\left(u_{b'}-u_{l}+2\eta\right)}{\sinh\left(u_{b'}-u_{l}-2\eta\right)}\frac{1}{\sinh\left(u_{b'}-v_{a}\right)}\prod_{j=1\neq a}^{N}\sinh\left(u_{b'}-v_{j}-2\eta\right)\\
-\frac{1}{\sinh\left(u_{b'}-v_{a}\right)}\prod_{j=1\neq a}^{N}\sinh\left(u_{b'}-v_{j}+2\eta\right)
\end{multline}
for $b'\neq b$
\begin{eqnarray}
\tilde{T}_{ab'}^{\left(b\right)} & = & \left(-1\right)^{N}\prod_{l=1\neq b'}^{N}\sinh\left(u_{b'}-u_{l}+2\eta\right)\prod_{j=1}^{N}\frac{\cosh\left(v_{j}-\eta\right)}{\cosh\left(u_{j}-\eta\right)}\frac{\sinh\left(2\eta\right)}{\cosh\left(v_{a}+\eta\right)\cosh\left(v_{a}-\eta\right)}\label{eq:btab}
\end{eqnarray}
for $b'=b.$ Note that $\tilde{T}_{ab'}^{\left(b\right)}=T_{ab'}$
for $b\neq b$.

Finally, the summation over $b$ in Eq.\@ (\ref{psid_psi_1_sum_of_dets})
can be evaluated using a general matrix identity: $\det\hat{T}+\sum_{b=1}^{N}\det\hat{\tilde{T}}^{\left(b\right)}=\det\left(\hat{T}+\hat{B}\right)$
where $\hat{\tilde{T}}^{\left(b\right)}$ is obtained from the matrix
$\hat{T}$ by substituting $b^{th}$column from matrix $\hat{B}$.
After constructing the matrix $\hat{B}$ out matrix elements $\tilde{T}_{ab}^{\left(b\right)}$
from Eq.\@ (\ref{eq:btab}) and performing the summation over $b$ in
Eq.\@ (\ref{psid_psi_1_sum_of_dets}), the form factor reads
\begin{equation}
\left\langle \mathbf{v}|\rho_{1}|\mathbf{u}\right\rangle =\prod_{j=1}^{N}\frac{\cosh\left(u_{j}-\eta\right)}{\cosh\left(u_{j}+\eta\right)}\frac{\sinh^{N}\left(2\eta\right)}{\prod_{j<i}\sinh\left(v_{j}-v_{i}\right)\prod_{j<i}\sinh\left(u_{j}-u_{i}\right)}\det\hat{K}\label{eq:psidpsi_ABA}
\end{equation}
where the matrix elements of $\hat{K}$ are 
\begin{eqnarray}
K_{ab} & = & \prod_{l=1\neq b}^{N}\frac{\sinh\left(u_{b}-u_{l}+2\eta\right)}{\sinh\left(u_{b}-u_{l}-2\eta\right)}\frac{1}{\sinh\left(u_{b}-v_{a}\right)}\prod_{j=1\neq a}^{N}\sinh\left(u_{b}-v_{j}-2\eta\right)-\frac{1}{\sinh\left(u_{b}-v_{a}\right)}\prod_{j=1\neq a}^{N}\sinh\left(u_{b}-v_{j}+2\eta\right)\nonumber \\
 &  & +\left(-1\right)^{N}\prod_{l=1\neq b}^{N}\sinh\left(u_{b}-u_{l}+2\eta\right)\prod_{j=1}^{N}\frac{\cosh\left(v_{j}-\eta\right)}{\cosh\left(u_{j}-\eta\right)}\frac{\sinh\left(2\eta\right)}{\cosh\left(v_{a}+\eta\right)\cosh\left(v_{a}-\eta\right)}\label{eq:Kab}
\end{eqnarray}

Now we evaluate the long wavelength limit for the result above. Applying
the inversion mapping from the algebraic to the coordinate representation
from Eq.\@ (\ref{eq:ABA_CBA_lwl}) to the matrix elements in Eq.\@ (\ref{eq:Kab})
and expanding the result up to the leading order $k_{j}^{u},k_{j}^{v}\ll1$
we obtain, 
\begin{equation}
K_{ab}=\left(-1\right)^{N-1}2mU\left(\left(mU\right)^{2}-1\right)^{\frac{N-2}{2}}\frac{\sum_{j=1}^{N}k_{j}^{v}-\sum_{j=1}^{N}k_{j}^{u}-k_{a}^{v}+k_{a}^{u}}{k_{b}^{u}-k_{a}^{v}}+2\left(mU+1\right)\left(\left(mU\right)^{2}-1\right)^{\frac{N-2}{2}}.
\end{equation}
Repeating the same procedure for the prefactor in Eq.\@ (\ref{eq:psidpsi_ABA})
and pulling a common scalar factor out of the matrix elements under
the determinant we obtain
\begin{equation}
\left\langle \mathbf{v}|\rho\left(0\right)|\mathbf{u}\right\rangle =\left(-1\right)^{N^{2}}2^{N}\frac{\left(mU\right)^{N}}{\left(mU+1\right)^{N}}\left(mU-1\right)^{N^{2}}\det\hat{\mathcal{K}}\label{eq:psidpsi_lwl}
\end{equation}
where the matrix elements of $\hat{\mathcal{K}}$ are 
\begin{equation}
\mathcal{K}_{ab}=\frac{k_{a}^{u}-k_{a}^{v}-\Delta P}{k_{b}^{u}-k_{a}^{v}}+\left(-1\right)^{N-1}\frac{mU+1}{mU},
\end{equation}
and $\Delta P=\sum_{j=1}^{N}k_{j}^{u}-\sum_{j=1}^{N}k_{j}^{v}$.

Evaluating the determinant we obtain
\begin{equation}
\det\mathcal{K}=\frac{mU+1}{mU}\frac{\left(\Delta P\right)^{N}\prod_{i<j}\left(k_{i}^{u}-k_{j}^{u}\right)\prod_{i<j}\left(k_{i}^{v}-k_{j}^{v}\right)}{\prod_{i,j}\left(k_{i}^{u}-k_{j}^{v}\right)}.
\end{equation}
This formula can be proved by induction analogously to the proof
of Eq.\@ (\ref{eq:cD}).

The form factor appearing in the dynamical structure factor in Eq.
(\ref{eq:S_continuum}) is a modulus squared of Eq.\@ (\ref{eq:psidpsi_lwl}).
Normalising the initial and the finial states using Eq.\@ (\ref{eq:norm_lwl})
as $\left|\left\langle f|\rho\left(0\right)|0\right\rangle \right|^{2}=\left|\left\langle \mathbf{k}^{f}|\rho\left(0\right)|\mathbf{k}^{0}\right\rangle \right|^{2}\left\langle \mathbf{k}^{f}|\mathbf{k}^{f}\right\rangle ^{-1}\left\langle \mathbf{k}^{0}|\mathbf{k}^{0}\right\rangle ^{-1}$
we obtain 
\begin{equation}
\left|\left\langle f|\rho\left(0\right)|0\right\rangle \right|^{2}=\frac{\left(mU\right)^{2N-2}}{\left(mU+1\right)^{2N-2}}\frac{P_{f}^{2N}\prod_{i<j}^{N}\left(k_{i}^{0}-k_{j}^{0}\right)^{2}\prod_{i<j}^{N}\left(k_{i}^{f}-k_{j}^{f}\right)^{2}}{\left(\mathcal{L}-\frac{NmU}{1+mU}\right)^{2N}\prod_{i,j=1}^{N}\left(k_{i}^{0}-k_{j}^{f}\right)^{2}}\label{eq:psidpsi_2}
\end{equation}
where $P_{0}=0$ for the ground state. Eq.\@ (\ref{eq:psidpsi_2}) is
Eq.\@ (\ref{eq:FF_dsf}) in the main part of the text with $Z=mU/\left(mU+1\right)/\left(\mathcal{L}-NmU/\left(1+mU\right)\right)$.

Note that when the final state becomes the ground state, $\mathbf{k}^{f}=\mathbf{k}^{0}$,
Eq.\@ (\ref{eq:psidpsi_2}) is divergent. In this case the matrix element
is evaluated using only translational symmetry and the definition
of the number of particles operator as
\begin{equation}
\left|\left\langle f|\rho\left(0\right)|0\right\rangle \right|^{2}=\frac{N^{2}}{\mathcal{L}^{2}},
\end{equation}
which also follows directly from Eqs.\@ (\ref{eq:psidpsi_ABA}, \ref{eq:Kab}),
by taking the limit $\mathbf{v}\rightarrow\mathbf{u}$. 
\end{document}